\documentclass[preprint]{aastex}

\usepackage{psfig}
\usepackage{graphicx}

\slugcomment{Accepted by \aj}

\shorttitle{Compact Groups in the SDSS}
\shortauthors{Lee et al.}

\begin{document}

\title{A Catalog of Compact Groups of Galaxies in the SDSS Commissioning Data}

\author{
Brian C. Lee,\altaffilmark{1}
Sahar S. Allam,\altaffilmark{3,2}
Douglas L. Tucker,\altaffilmark{2}
James Annis,\altaffilmark{2}
Michael R. Blanton,\altaffilmark{4}
David E. Johnston,\altaffilmark{5,2}
Ryan Scranton,\altaffilmark{6,5}
Yamina Acebo,\altaffilmark{2,7}
Neta A. Bahcall,\altaffilmark{8}
Matthias Bartelmann,\altaffilmark{9}
Hans B\"{o}hringer,\altaffilmark{10}
Nancy Ellman,\altaffilmark{11}
Eva K. Grebel,\altaffilmark{12}
Leopoldo Infante,\altaffilmark{8,13}
Jon Loveday,\altaffilmark{14}
Timothy A. McKay,\altaffilmark{7}
Francisco Prada,\altaffilmark{12}
Donald P. Schneider,\altaffilmark{15}
Chris Stoughton,\altaffilmark{2}
Alexander S. Szalay,\altaffilmark{16}
Michael S. Vogeley,\altaffilmark{17}
Wolfgang Voges,\altaffilmark{10}
and Brian Yanny\altaffilmark{2}
}

\altaffiltext{1}{Lawrence Berkeley National Laboratory, 1 Cyclotron Rd, Berkeley CA 94720-8160}
\altaffiltext{2}{Fermi National Accelerator Laboratory, P.O. Box 500, Batavia, IL 60510.}
\altaffiltext{3}{Department of Astronomy, New Mexico State University, 1320 Frenger Mall, Las Cruces, NM 88003-8001.}
\altaffiltext{4}{Department of Physics, New York University, 4 Washington Place, New York, NY 10003.}
\altaffiltext{5}{Department of Astronomy and Astrophysics, The University of Chicago, 5640 South Ellis Avenue, Chicago, IL 60637.}
\altaffiltext{6}{Department of Physics \& Astronomy, University of Pittsburgh, 3941 O'Hara Street, Pittsburgh, PA 15260 USA.}
\altaffiltext{7}{Department of Physics, University of Michigan, 500 East University, Ann Arbor, MI 48109-1120.}
\altaffiltext{8}{Princeton University Observatory, Peyton Hall, Princeton, NJ 08544.}
\altaffiltext{9}{Max-Planck-Institut f\"{u}r Astrophysik, Postfach 1317, D-85741 Garching, Germany.}
\altaffiltext{10}{Max-Planck-Institut f\"{u}r extraterrestische Physik, Postfach 1312, D-85741 Garching, Germany.}
\altaffiltext{11}{Department of Physics, Yale University, PO Box 208121, New Haven, CT 06520.}
\altaffiltext{12}{Max-Planck-Institut f\"{u}r Astronomie, K\"{o}nigstuhl 17, D-69117 Heidelberg, Germany.}
\altaffiltext{13}{Pontificia Universidad Cat\'{o}lica de Chile, Departamento de Astronom\'{\i}a y Astrof\'{\i}sica, Facultad de F\'{\i}sica, Casilla 306, Santiago 22, Chile}
\altaffiltext{14}{Astronomy Centre, University of Sussex, Falmer, Brighton BN1 9QJ, UK.}
\altaffiltext{15}{Department of Astronomy and Astrophysics, The Pennsylvania State University, 525 Davey Laboratory, University Park, PA 16802.}
\altaffiltext{16}{Department of Physics and Astronomy, The Johns Hopkins University, 3400 North Charles Street, Baltimore, MD 21218-2686.}
\altaffiltext{17}{Department of Physics, Drexel University, 3141 Chestnut Street, Philadelphia, PA 19104.}


\begin{abstract}

Compact groups (CGs) of     galaxies --- relatively poor  groups    of
galaxies in which  the typical separations  between members  is of the
order of a galaxy diameter --- offer an exceptional laboratory for the
study of dense galaxian environments with  short ($<$ 1~Gyr) dynamical
time-scales.

\noindent 
In this paper, we  present  an objectively defined  catalog of  CGs in
153~sq~deg of  the Sloan   Digital   Sky Survey Early  Data    Release
(SDSS~EDR).   To  identify  CGs, we  applied    a modified version  of
Hickson's  \citeyearpar{Hickson82}  criteria    aimed  at finding  the
highest density CGs and thus reducing the number of chance alignments.
Our catalog contains  175 CGs   down to a  limiting
galaxy  magnitude of $r^* = 21$.   The resulting catalog  has a median
depth of $z_{\rm   med}   \approx 0.13$,   substantially  deeper  than
previous CG catalogs.  Since the SDSS  will eventually image up to one
quarter of the  celestial sphere, we  expect our  final catalog, based
upon the completed SDSS, will contain on the  order of 5,000 -- 10,000
CGs.    This catalog will be   useful  for conducting  studies of  the
general   characteristics  of  CGs,    their environments,  and  their
component galaxies.
\end{abstract}


\section{Introduction}
\label{intro}

Perhaps over half of all galaxies lie within groups containing 3 -- 20
members \citep{Tully87}; yet, due to the difficulty of discerning them
from  the field, groups  of  galaxies are, as  a   whole, not as  well
studied as  larger galaxy systems.   Compact groups of galaxies (CGs),
however,  defined by  their small  number  of members ($<$ 10),  their
compactness (typical intra-group  separations of a galaxy  diameter or
less), and   their relative  isolation (intra-group separations  $\ll$
group-field  separations) are  readily  identifiable.   Studies of CGs
mainly concentrate on two issues: (1) What is  the origin and relative
importance of CGs in  the Universe?  (2)  Is there a relation  between
the global properties of these systems and the formation and evolution
of their member  galaxies?  The possibility that  these two issues are
connected   makes CGs  particularly interesting  (see   the  review by
\citealt{Hickson97}).

The first example of a CG was found over one hundred years ago by
\citet{Stephan77}. The best known   catalog  is that of  the   Hickson
Compact Groups (HCGs; \citealt{Hickson82,Hickson93}), a sample
comprising 100 groups selected from the red (E) prints of the Palomar
Observatory Sky Survey (POSS).  Other catalogs now available include
an initial DPOSS CG catalog \citep{Iovino03}, the Southern CG
catalog \citep{Iovino02,Prandoni94}, and those extracted from the 3D
UZC galaxy catalog \citep{FocardiKelm02} and from the CfA2
\citep{Barton96} and    Las  Campanas \citep{AllamTucker00}   redshift
surveys.

Due  to their high densities  (equivalent to those   at the centers of
rich clusters) and low velocity dispersions (roughly 200~km~s$^{-1}$),
CGs  represent  an environment  where interactions,  tidally triggered
activity, and galaxy mergers are expected to be more prevalent than in
most other environments.  Studies of interacting galaxy pairs, both in
the infrared and ultraviolet, suggest that interactions can trigger an
inflow of gas to  the galactic nucleus,  resulting in either starburst
or  AGN activity.  Although individual     CGs are known to    contain
starbursts and  AGN \citep{Menon95,Ribeiro96}, samples  of CG galaxies
as a whole do  not appear to   show rates of either activity  enhanced
beyond that of the field \citep{Pildis95,Allam98,Allam99}.

Some calculations have  predicted extremely short  dynamical lifetimes
for CGs, $t_{\rm  dyn} <$ 1~Gyr, leading  to speculation that many  or
most CGs may in  fact be chance alignments  instead of real structures
\citep{Mamon86}.  However, evidence of  interactions and the discovery
of diffuse intra-group X-ray gas within perhaps as many as 75\% of the
HCGs \citep{Hickson97,Mulchaey00} attest to their physical reality.

N-body simulations have pointed  out two possible classes of  solution
to the short   dynamical time problem.  The  first  is that  there  is
ongoing formation of CGs, and  the longevity of   the group is due  to
secondary infall.  In this  case, CGs must  be continually forming  in
moderately   dense environments like those  of   loose groups, and, in
fact,     many   are seen   to     be   embedded   in    such  systems
\citep{Ramella94,Barton96,Diaf94,Diaf95,deCarvalho97,Coziol98,Ribeiro98,Tovmassian01,Tovmassian02}.

In the second class of solution, the longevity of CGs is due to either
their  specific initial conditions or  a massive halo encompassing the
entire   group \citep{Athanassoula00}. \citet{Athanassoula97} have shown
that CGs with an appropriate  arrangement of luminous and dark  matter
can persist  for $\sim$ 25~Gyr; that  massive common dark matter halos
around CGs might  indeed   exist is not  inconsistent  with   with the
analysis of the dynamics of satellite galaxies within individual CGs
\citep{Perea00}.

Thus, CGs have  sparked intense interest,  both in their formation and
evolutionary  histories, and in  the interaction  phenomena associated
with these dense environments.   Our difficulties in understanding the
existence of CGs and their dynamics may be  a consequence of the small
samples studied so   far.  Further  progress  will  require a  larger,
deeper,  and more uniform  sample of   CGs with which  to study  their
nature,  evolution, and     origin.   The Sloan   Digital Sky   Survey
\citep{York00}, which  will eventually cover up  to one quarter of the
sky with uniform  photometry in five filters, makes  for an obvious CG
hunting ground.

We have thus  embarked on a project  to extract an objectively defined
catalog of CGs from the SDSS \citep{Lee01a,Lee01b}, starting with Runs
752 and 756 of the SDSS  Early Data Release (EDR) \citep{Stoughton02}.
We present  this initial   catalog  as follows.  In   \S~\ref{data} we
describe the region of the sky used for this preliminary search.  In
\S~\ref{cat_construct}, we   describe   our CG    catalog construction
techniques.  In \S~\ref{catalog} we present  the catalog and an  atlas
of corresponding postage-stamp images for the individual groups.  In
\S~\ref{comparison_other} we compare the properties of this catalog to
those of  previous CG catalogs.   In \S~\ref{morph-env} we compare the
properties of SDSS CG member galaxies to  the properties of SDSS field
galaxies. In \S~\ref{conclusions} we conclude  and describe our future
plans.  Throughout the paper, a flat $\Lambda$ cosmological model with
$\Omega_{\rm  M} =   0.3$,  $\Omega_{\Lambda}  =  0.7$,  and  $H_0   =
100h$~km~s$^{-1}$~Mpc$^{-1}$ is assumed.


\section{The Data}
\label{data}
 
The   SDSS has  as  its goal   to  image  up  to 10,000~sq~deg  ($\pi$
steradians)  of  the Northern Galactic  Cap  in five different filters
($u$,  $g$, $r$,  $i$, $z$) and  to  perform followup spectroscopy  of
$10^6$  galaxies and $10^5$ quasars selected  from the imaging survey.
The Survey will be complete (S:N  $\sim$ 5:1) to  limiting $u$ $g$ $r$
$i$  $z$ magnitudes  of   roughly  22.3,  23.3,  23.1,  22.3  and 20.8
respectively.   An additional  three stripes in  the Southern Galactic
Cap will be observed repeatedly for variability studies and will reach
a co-added depth of $g \sim 25$.

While the SDSS  was originally planned to  be on the  $u' g' r' i' z'$
system of \citet{Fukugita96}, the delivered  filters do not  perfectly
conform to this  system.  The photometric  calibration system for SDSS
(called $u  g r i  z$) had only been  finalized as  of the recent Data
Release 1 (DR1) \citep{Abazajian03} , so the preliminary magnitudes as
presented in this  paper, which are based  upon a pre-DR1 calibration,
will be denoted as $u^*$,  $g^*$,  $r^*$, $i^*$,  and $z^*$ (for  more
details  see \citealt{Stoughton02}).  The  system  used in this  paper
should differ absolutely from the final ($u g r i z$) SDSS photometric
system  by    only   a    few  percent.   [See     \citet{Fukugita96},
\citet{Gunn98},    \citet{York00}, \citet{Hogg01},   \citet{Lupton01},
\citet{Smith02},      \citet{Stoughton02},   \citet{Blanton03a},   and
\citet{Pier03}  for   further  details   about  the   photometric  and
spectroscopic surveys.]

The base galaxy catalog  for the present study  was generated from the
spring     equatorial    scan subset      of   the  SDSS    EDR   data
\citep{Stoughton02}.  Instead of using the star-galaxy classifications
from the   standard SDSS imaging pipeline   ({\tt photo}), however, we
chose to use the  more robust classifications measured  by the code of
\citet{Scranton02}.    Galaxies were  separated  from  stars  using  a
Bayesian method which combines the  automated classifications from the
photometric  pipelines with a  correction for variations in seeing and
assigns     a  probability  for   each    object    being a     galaxy
\citep{Scranton02}.  The accurate astrometry
\citep{Pier03}  combined  with  color  and morphological   information
derived  from the SDSS   digital  images \citep{Lupton01} allows   for
robust   star-galaxy separation to a    limiting magnitude of $\sim22$
(with   1.75 arcsecond or  better  seeing).  Very few  objects have an
uncertain star-galaxy probability (near 50\%); thus we have simply cut
at 50\%.  This  method  suffers less  at the  faint end  from  stellar
contamination than  does the standard SDSS~EDR star-galaxy separation.
Our  earliest CG catalogs  using pre-EDR versions  of the SDSS outputs
had stellar contamination  (a star identified as  a galaxy) in as many
as 25\% of the selected CGs.  Using the \citet{Scranton02} catalog and
cutting at a galaxy likelihood of 50\%, our catalogs now have at worst
$5\%$ stellar contamination.

This star-galaxy separation breaks down for small, faint galaxies, and
routines used for  this initial study  do not measure  bright or large
galaxies well.   The combination limits  the present study to galaxies
within the $r^*$ magnitude  range of  14.0 to  21.0 (which limits  the
brightest   CG  member to      $r^*$   between 14.0 and    18.0,   see
\S~\ref{cat_construct}), excluding much of  the HCG catalog but  still
providing some overlap.

\begin{figure*}
\centering
\includegraphics[angle=0,scale=1.0]{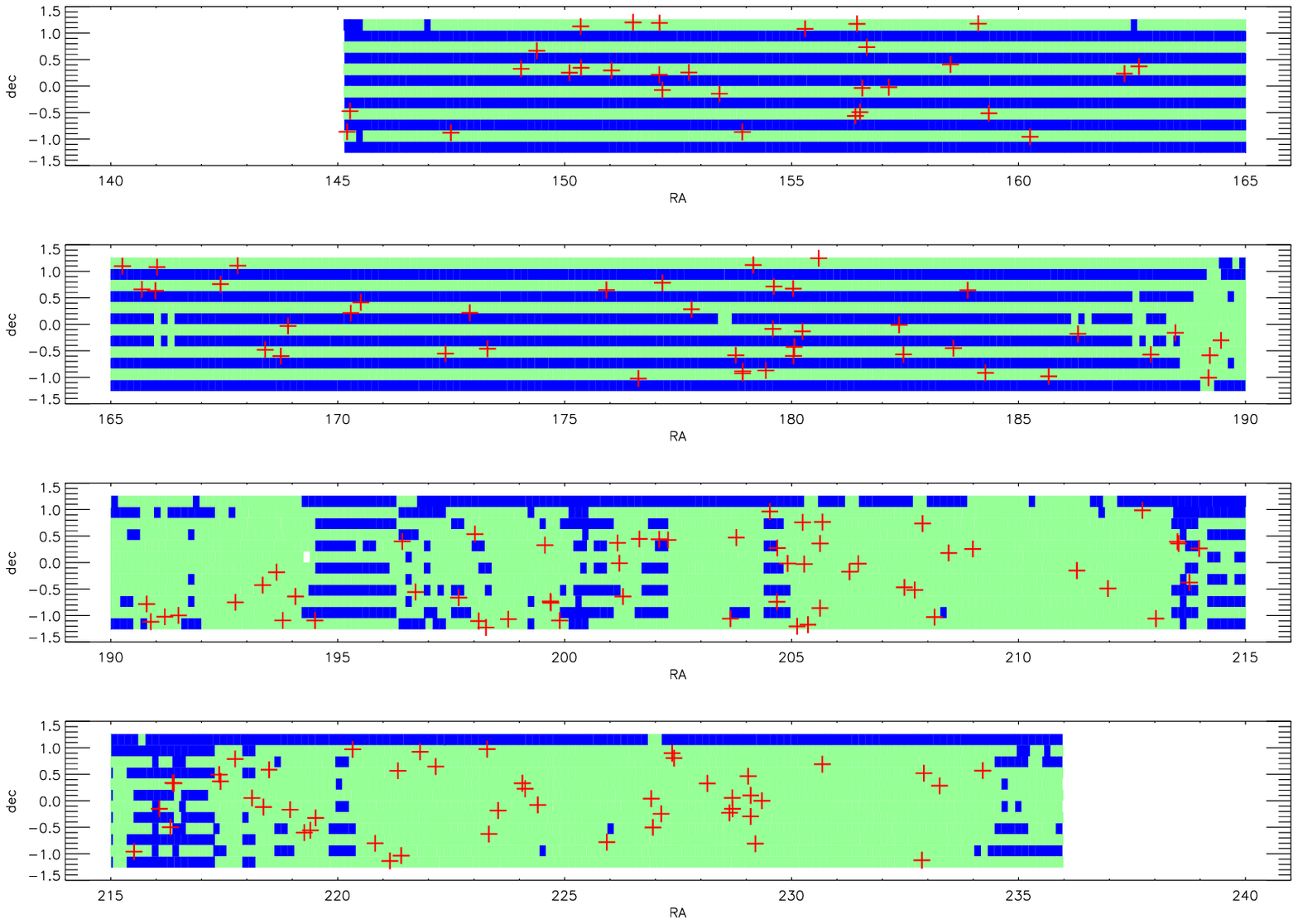}
\caption{The 230~sq~deg area on the sky used for the present 
study, from the \citet{Scranton02} catalog of Runs 752  and 756.  {\em
Green (light) blocks\/} are individual  fields for which the seeing is
$\leq 1.6$ arcsec (153~sq~deg total);  {\em blue (dark) blocks\/} are
fields  for which the  seeing is  $>$ 1.6 arcsec.   Galaxies in  these
fields were  used to test   isolation but compact groups  with  member
galaxies  in these fields  have been excluded  from our catalog.  {\em
Red crosses\/} denote the locations of the 175 CGs extracted from this
area. \label{sky_dist}}
\end{figure*}

The data  presented   here   are from Runs    752  and  756  (a  $\sim
2.5^{\circ}$ wide stripe  from $\alpha = 145^{\circ}$ to $236^{\circ}$
centered on $\delta$ = $0^{\circ}$)  which were observed on the nights
of 1999 March  20 and 22.  Galaxies  from the entire $\sim$ 230~sq~deg
region were used  for checking  the  isolation  criteria described  in
\S~\ref{cat_construct}, however, in order  to obtain the least stellar
contamination of the galaxy sample,  only CGs observed in frames  with
seeing  $\leq 1.6$ arcseconds were retained  for the catalog, yielding
an effective area  on the sky for our  base sample  of 153~sq~deg (see
Fig.~\ref{sky_dist}).

Finally, we note that the  \citet{Scranton02} galaxy catalog was based
upon an earlier processing (``rerun'')  of the imaging data reductions
than  is presented    in    the  final   SDSS~EDR.   As   a    result,
$u^*g^*r^*i^*z^*$  magnitudes may   differ  slightly from the   values
obtained by  a   direct query of  the SDSS~EDR   database.  Similarly,
although each object's run,   camera column, and  field identification
numbers will remain the same,  the object identification number within
the field  will likely differ from  the value provided by the SDSS~EDR
database.


\section{Catalog Construction}
\label{cat_construct}

Hickson's \citeyearpar{Hickson82} criteria for extracting compact
groups from the POSS red (E) prints include the following:

\begin{itemize}
\item   $N \ge 4$                         (population), 
\item   $\theta_N \ge 3 \times \theta_G$  (isolation), and 
\item   $\mu_G < 26.0$~mag~arcsec$^{-2}$  (compactness), 
\end{itemize}

\noindent where 

\begin{itemize}
\item   $N$ is the total number of galaxies within 3 magnitudes of the
	brightest group member,
\item   $\mu_G$ is the total magnitude   of these galaxies per  square
	arcsec  averaged  over  the smallest circle  containing  their
	geometric centers (note that using  a mean surface  brightness
	yields, to   first  order, a distance-independent   measure of
	compactness),
\item   $\theta_G$ is the angular diameter of this smallest circle, and 
\item   $\theta_N$  is the angular diameter  of the largest concentric
	circle that contains no other  (external) galaxies within this
	magnitude range or brighter.
\end{itemize}

\noindent Searching the POSS prints, which cover  67\% of the celestial 
sphere,  by  eye, Hickson  found 100 CGs;  the median  redshift of this
sample is $z_{\rm med} = 0.03$.

To  select our SDSS  CGs we   use computer  code embodying  a slightly
modified  Hickson criteria to  extract  a CG  catalog from this galaxy
catalog:

\begin{itemize}
\item    $14.0 \le r^* \le 21.0$, 
\item    $10 \ge N \ge 4$, 
\item    $\theta_N > 3 \times \theta_G$, and  
\item    $\mu_G < 24.0$~mag~arcsec$^{-2}$ (in  SDSS $r^*$ band), 
\end{itemize}
where $r^*$ is the SDSS  $r$-band {\em model\/} magnitude, which tends
to  be more robust than  the SDSS $r$-band {\em Petrosian\/} magnitude
for galaxies fainter than $r^* \sim 18$ (see \citealt{Stoughton02} for
definitions of the various SDSS magnitudes).

Three changes were made from the Hickson criteria.  The first merely
added an upper limit of $N \le 10$ to the number of members in a CG.
This additional limit was added in order to constrain the run-time of
the group-finding algorithm, which is basically an $n^2$ process.
Since the most populous group in our catalog contains only $N=7$
members, this was (intentionally) not a very restrictive constraint.
The second modification simply requires that $\theta_N$ is strictly
greater than $3 \times \theta_G$ instead of greater than or equal to;
this change simplified the coding of the algorithm which identifies
isolated groups and finds the smallest enclosing circle.

\begin{figure*}
\centering
\includegraphics[angle=90,scale=0.6]{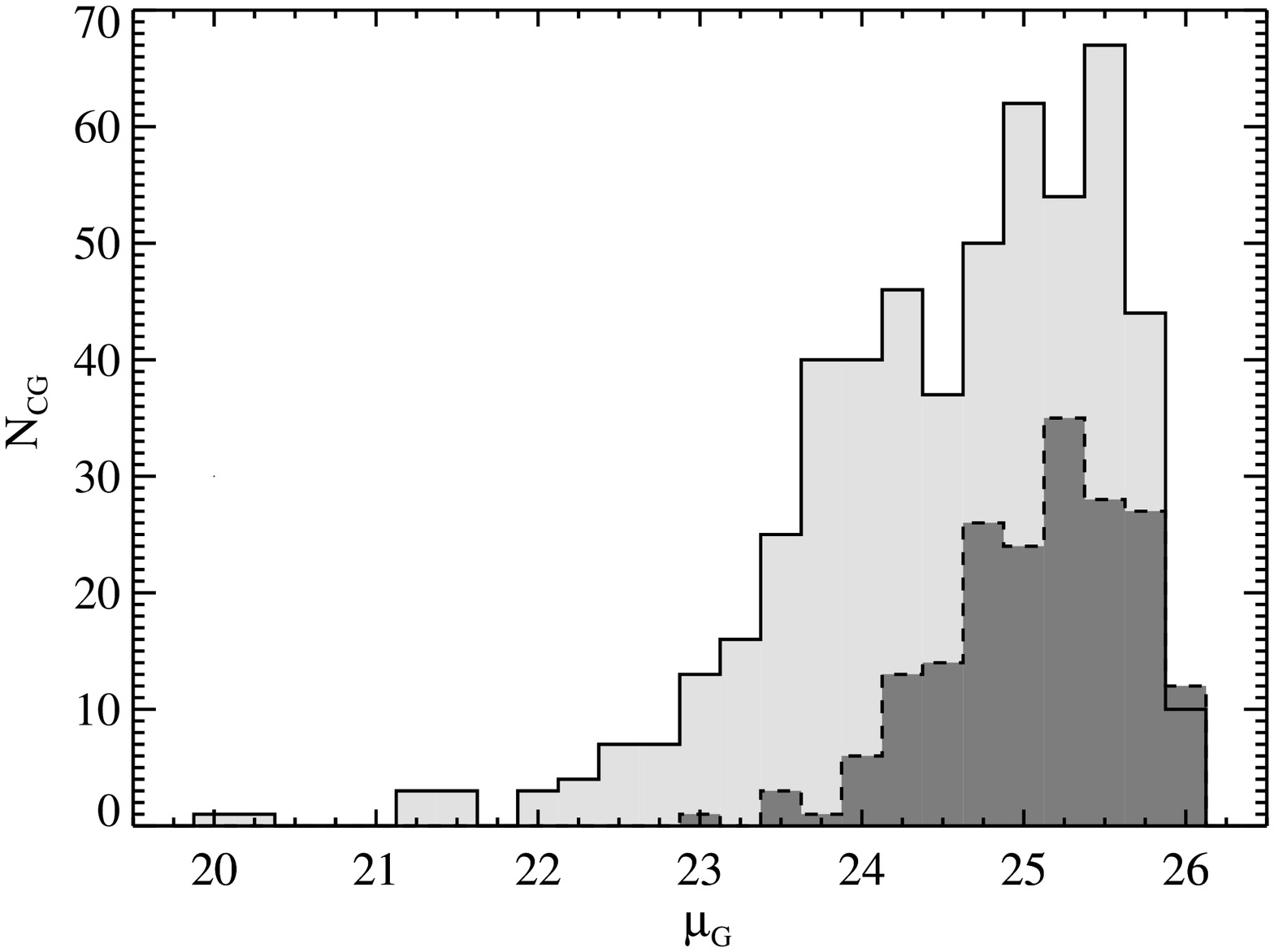}
\includegraphics[angle=90,scale=0.6]{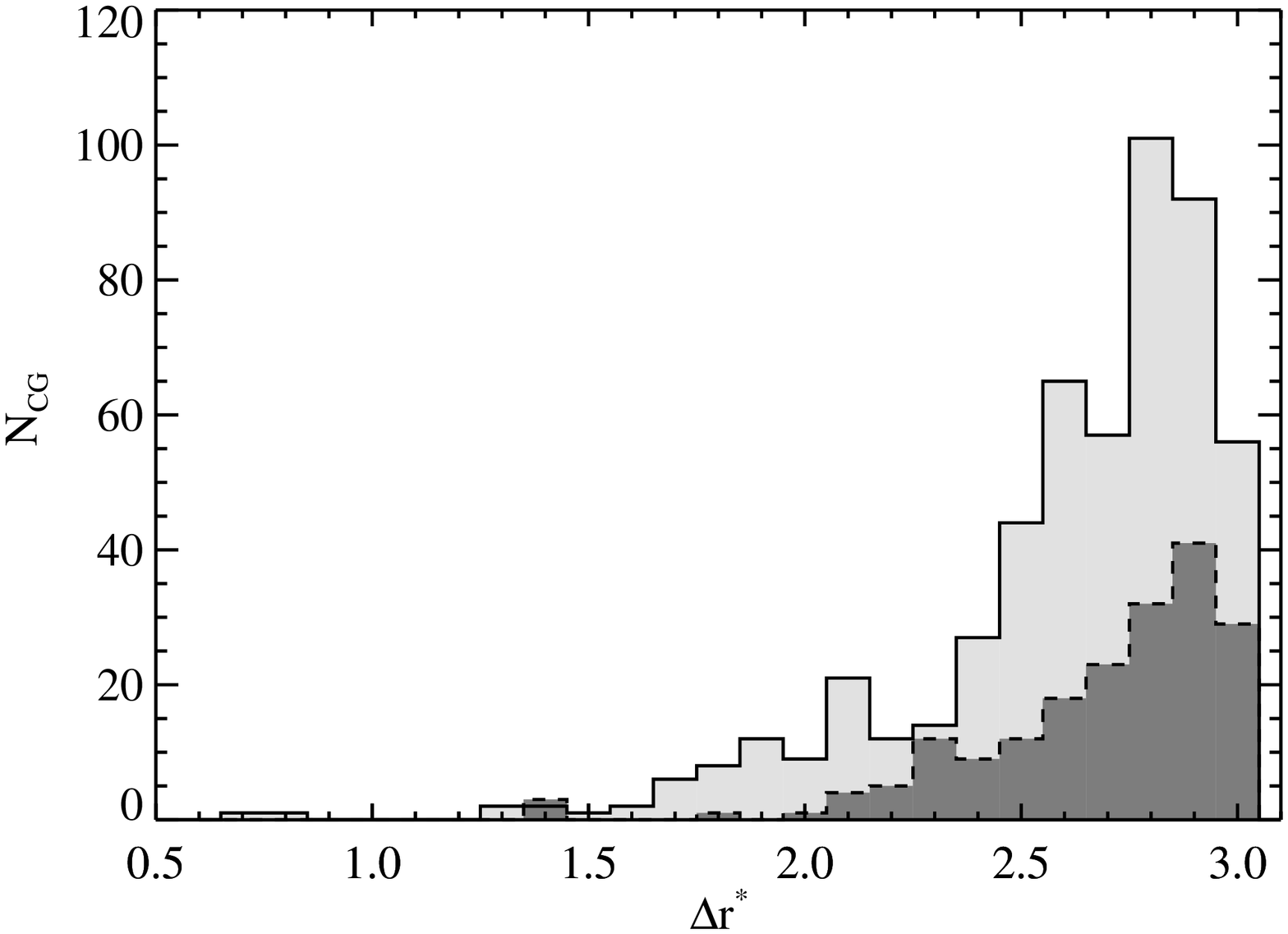}
\caption{Number of compact groups found in the real (light/solid line)
and randomized (dark/dashed line) galaxy catalogs. (a) Histogram of the surface
brightness.  (b) Histogram of the difference between the dimmest and
the bright group members, $\Delta r^*$.  
\label{fig1}}
\end{figure*}

The third, the change from $\mu_{G} < 26.0$ to $\mu_{G} < 24.0
$~mag~arcsec$^{-2}$, resulted from tests suggested by 
Iovino (private communication, see also \citealt{Iovino01, Iovino02,
Iovino03}) to reduce the rate of false CG detection.  Following the
approach described by \citealt{Iovino02} and \citealt{Iovino03}, we
estimated the number of non-real CGs produced through random
alignments as a function of $\mu_{G}$,
and compared it with the number of CGs found in the original SDSS
catalog.  Using a subset of our galaxy catalog, we generated a
randomized galaxy catalog by assigning to each galaxy an RA and DEC
drawn at random from within the area of the subset; a value for the
seeing at this random position was taken to be that for the nearest
galaxy to this position in the real catalog.  The compact group
finding code was then run on this randomized galaxy catalog to
generate a list of false CGs produced by projection effects, and this
fake CG catalog was compared to the portion of our real CG catalog
from the same region.  In fig (\ref{fig1},a) shows the histogram of
surface brightness for CGs in each catalog.  The ratio of random to
real CGs is $\approx$ 4\% when restricting the search to CGs with
$\mu_{G} <$ 24.0, while the ratio increases to 55\% within 24.0$<$
$\mu_{G} \leq$ 25.0 and 80\% within 25.0$<$ $\mu_{G} \leq$ 26.0.  From
this plot it is clear that cutting at $\mu_{G} <$ 24 leaves a much
more clean (and compact) catalog.  In fig (\ref{fig1},b) we show a
histogram of the difference in magnitude between the dimmest and
brightest group member ($\Delta r^{*}$) where the maximum allowed by
the Hickson criteria is 3.  In contrast with Iovino et al. (2003), for
our sample we found no strong trend in the contamination rate as a
function $\Delta r^{*}$.  There may be a small offset between the
peaks, but a large gain in detection efficiency was unlikely to come
from modifying this selection parameter in our algorithm.  We
therefore restricted the surface brightness limit to be $\mu_{G} < $
24 ~mag~arcsec$^{-2}$, and left $\Delta r^{*}$ unchanged, in order to
reduce the contamination rate of our sample to reasonable values.

In addition, we examined the isolation criterion, which required that
the next closest galaxy bright enough to be in the group or brighter
be at least three group radii away.  Allam \& Tucker (2000) found that
one third of their compact groups are part of larger structures, also
de Carvalho et al. (1997) and Ribeiro et al.  (1998) found that a
large fraction of HCGs are part of larger structures, and only a small
number of groups can be defined as truly isolated.  To overcome this
problem, Iovino et al. (2003) adopt a flexible isolation criterion,
such that no galaxy of magnitude brighter than mag of the faintest CG
member + 0.5 mag is within the isolation radius.  Our algorithm, which
builds groups based on adding the closest galaxies which are in the
correct magnitude range to be members, did not allow any flexibility
in lowering the isolation criteria.  However we did note that the
majority of the closest non-member galaxies (which were eligible for
membership) for our groups were within $\pm$0.5 magnitudes of the
dimmest group member (73\%) and were between three and four group
radii of the center of the group (51\%).  (Numbers for the random
catalog were similar, at 73\% within $\pm$0.5 magnitudes and 40\%
nearby, thus relaxing isolation will add random alignments in nearly
equal proportion to real groups.)  Relaxing the isolation criteria for
dim galaxies should be explored in our future efforts, as it seems
likely the strict isolation criteria used excluded some compact groups.

Using our modified Hickson criteria, we found a total of 232 candidate
CGs in 153~sq~deg.  In order to identify any potential weaknesses of
our selection technique and improve future versions, each of these
candidate CGs was inspected by eye, and much of the SDSS $r$ band
imaging data (the corrected frames) for these candidate CGs were
re-analyzed using GAIA \citep{GAIA} and SExtractor \citep{SExtractor}
to double-check galaxy identification.

Of these 232 candidate CGs, 175 survived inspection to be included in
the final CG catalog.  Of the 57 which were discarded, 55 were found
to be false detections due to poor deblending
\footnote{{\tt http://www.sdss.org/dr1/algorithms/deblend.html}}          
of saturated bright   stars  or angularly large  galaxies  and/or poor
magnitude determination of faint diffuse  objects in the original SDSS
image processing.  Another 2 candidate  groups were found to contain a
QSO   which was
mis-classified by  {\tt photo} and  SExtractor  as a  galaxy.  These 2
groups are listed as Tri since they contain only 3 galaxies.


\section{The Catalog}
\label{catalog}

Our final catalog  of SDSS CGs, extracted  from 153~sq~deg of sky from 
Runs 752 Run 756 of the \citet{Scranton02} galaxy catalogs in the SDSS
EDR, contains   a total of 744  galaxies  in 175 CGs.   Of  the 744 CG
member  galaxies, 158 have spectroscopic  redshifts (135 from the SDSS
itself).   Despite this low number, 131  (75\%) of SDSS CGs contain at
least one group member  with a spectroscopic redshift.  The properties
of the CGs are described in  Table~\ref{tab_cg}; the properties of the
CG member galaxies are described in Table~\ref{tab_cgmem}.

\subsection{The Properties of the Compact Groups}

In Table~\ref{tab_cg} we list the general properties of all 175
SDSSCGs (plus two Triplet systems):

\begin{itemize}

\item [Column  (1):] A running identification number.

\item [Column  (2):] $\alpha_{\rm circ}$, the RA (J2000.0) of the center
		     of    the smallest  circle  containing all  group
		     members   within  3 magnitudes  of  the brightest
		     group member.

\item [Column  (3):] $\delta_{\rm circ}$, the DEC (J2000.0) of the center 
		     of  the  smallest  circle   containing  all group
		     members  within  3  magnitudes of  the  brightest
		     group member.

\item [Column  (4):] $\theta_G$, the angular diameter of this smallest circle 
		     (in arcsec).

\item [Column  (5):] $\mu_G$, the total $r^*$ magnitude of the $N$ [Column (6)]
		     group members averaged  over  the smallest circle
		     containing their geometric centers.

\item [Column  (6):] $N$, the total number of galaxies in the group within 3
		     magnitudes of the brightest group member.

\item [Column  (7):] $\Delta r^*$, the difference in the $r^*$ magnitude between
		     the brightest group member and the faintest.

\item [Column  (8):] $\theta_N / \theta_G$, the ratio of $\theta_N$, the angular 
		     diameter  of  the largest  concentric circle that
		     contains  no  other (external)  galaxies within 3
		     magnitudes   of  the  brightest group member,  to
		     $\theta_G$, the  angular diameter of the smallest
		     circle containing  all group  members within this
		     magnitude range [Column (4)].

\item [Column  (9):] $z^{\rm bgm}_{\rm photo}$, the photometric redshift 
		     of the brightest   group member as determined  by
		     the method described in \S~\ref{redshift_dist}.

\item [Column (10):] $z^{\rm cg}_{\rm spectro}$, the spectroscopic redshift 
		     for   the  group based upon    the  median of the
		     spectroscopic redshifts  of  its  member galaxies
		     (see  Table~\ref{tab_cgmem}).    If  available, a
		     group member's  SDSS spectroscopic redshift  was
		     used; otherwise,  if available, its spectroscopic
		     redshift from NED (the  NASA  Extragalactic  Database), 
                     was used.

\item [Column (11):] $N^{\rm cg}_{\rm spectro}$, the number of group 
		     member  spectroscopic        redshifts   used  in
		     determining $z^{\rm cg}_{\rm spectro}$.

\item [Column (12):] comments
	\begin{itemize}

	\item [+1:]  a faint background galaxy is visible which could be a member of the group

	\item [+2:]  two faint background galaxies are visible which could be members 
			of the group 

	\item [isolated?:]  faint background galaxies are visible which could be members of the group

	\item [M(ab):] galaxies a and b appear to be merging

	\item [I(ab):] galaxies a and b appear to be interacting

	\item [T(a):] galaxy a shows a tidal tail

	\item [h(abc):] galaxies a, b, and c appear to be embedded within a common halo
	 	
	\item [QSO(a):] galaxy a is classified as a QSO in NED

	\end{itemize}

\end{itemize}

\subsection{The Properties of Group Members}

In Table~\ref{tab_cgmem} we list the general properties of the individual
member galaxies in each of the 175 SDSSCGs (and the 2 Triplet systems):

\begin{itemize}

\item [Column  (1):] Name of the group member, composed of the CG identification 
		     number  (from Column (1) of   Table~\ref{tab_cg})
		     and   an identification letter  for galaxy (where
		     ``a'' is  the brightest  group member,  ``b'' the
		     second brightest, ...).

\item [Column  (2):] $\alpha$, the RA (J2000.0) of the galaxy.

\item [Column  (3):] $\delta$, the DEC (J2000.0) of the galaxy.

\item [Column  (4):] The SDSS $r$-band model magnitude $r^{*}$ (non-de-reddened).

\item [Column  (5):] The estimated rms error in the SDSS $r$-band model magnitude 
			$\sigma_{r^{*}}$.

\item [Column  (6):] The reddening in $r^{*}$, $A(r^*) = 2.751 \times E(B-V)$, 
			as estimated from the \citet{Schlegel98} reddening maps.

\item [Column  (7):] $u^*-g^*$, where $u^*$ and $g^*$ are, respectively, the 
			 (non-de-reddened) SDSS $u$-band and $g$-band model magnitudes.

\item [Column  (8):] $\sigma_{u^*-g^*}$, the rms error in $u^*-g^*$ [Column (8)]
			estimated by adding the rms errors in $u^*$ and $g^*$ in quadrature.

\item [Column  (9):] $g^*-r^*$, where $g^*$ and $r^*$ are, respectively, the 
			(non-de-reddened) SDSS $g$-band and $r$-band model magnitudes.

\item [Column (10):] $\sigma_{g^*-r^*}$ , the rms error in $g^*-r^*$ [Column (10)]
			estimated by adding the rms errors in $g^*$ and $r^*$ in quadrature.

\item [Column (11):] $r^*-i^*$, where $r^*$ and $i^*$ are, respectively, the 
			(non-de-reddened) SDSS $r$-band and $i$-band model magnitudes.

\item [Column (12):] $\sigma_{r^*-i^*}$ , the rms error in $r^*-i^*$ [Column (12)]
			estimated by adding the rms errors in $r^*$ and $i^*$ in quadrature.

\item [Column (13):] $i^*-z^*$, where $i^*$ and $z^*$ are, respectively, the 
			(non-de-reddened) SDSS $i$-band and $z$-band model magnitudes.

\item [Column (14):] $\sigma_{i^*-z^*}$ , the rms error in $i^*-z^*$ [Column (14)]
			estimated by adding the rms errors in $i^*$ and $z^*$ in quadrature.

\item [Column (15):] $z_{\rm sp}$, the spectroscopic redshift for the galaxy as measured 
			by the SDSS, if available.  Of the 744 CG members, 135 have
			SDSS spectroscopic redshifts.  

\item [Column (16):] $z_{\rm NED}$, the spectroscopic redshift for the galaxy from 
			the   NASA  Extragalactic  Database (NED),  if
			available.   Of  the 744  CG members,  49 have
			spectroscopic redshifts  from NED.   Of  these
			49,    23  have no corresponding spectroscopic
			redshift from the SDSS.

\item [Column (17):] the source name of the NED-derived spectroscopic redshift [Column (16)].

\end{itemize}

\subsection{The Atlas}
\label{atlas}
 
For each  of the SDSSCGs  (and the  two Triplets) we  have prepared an
atlas image (see  Fig~\ref{fig:cg}).    These atlas images   have  the
following properties:

\begin{itemize}
\item the size of each image is 3$\times$ the group radius; 
\item the images were obtained from the $r$-band SDSS corrected frames; 
\item each image is labeled by the name of the group, which
	has the format ``SDSS0~CG$NN$'', where ``SDSS0'' refers to the
	fact that this catalog is based on SDSS commissioning data 
	(``Data Release 0''), and where $NN$ refers to the running ID 
	number of the group;
\item each group member is identified by a ``$\times$'' mark  and labeled 
      by its ID letter (a, b, c, d, ...);
\item the center of the group is marked by a square box; and 
\item a circle of diameter $\theta_G$ (as defined in \S~\ref{cat_construct}) 
	is also drawn.
\end{itemize}

\subsection{The Redshift Distribution}
\label{redshift_dist}

\begin{figure*}
\centering
\includegraphics[angle=0,scale=0.8]{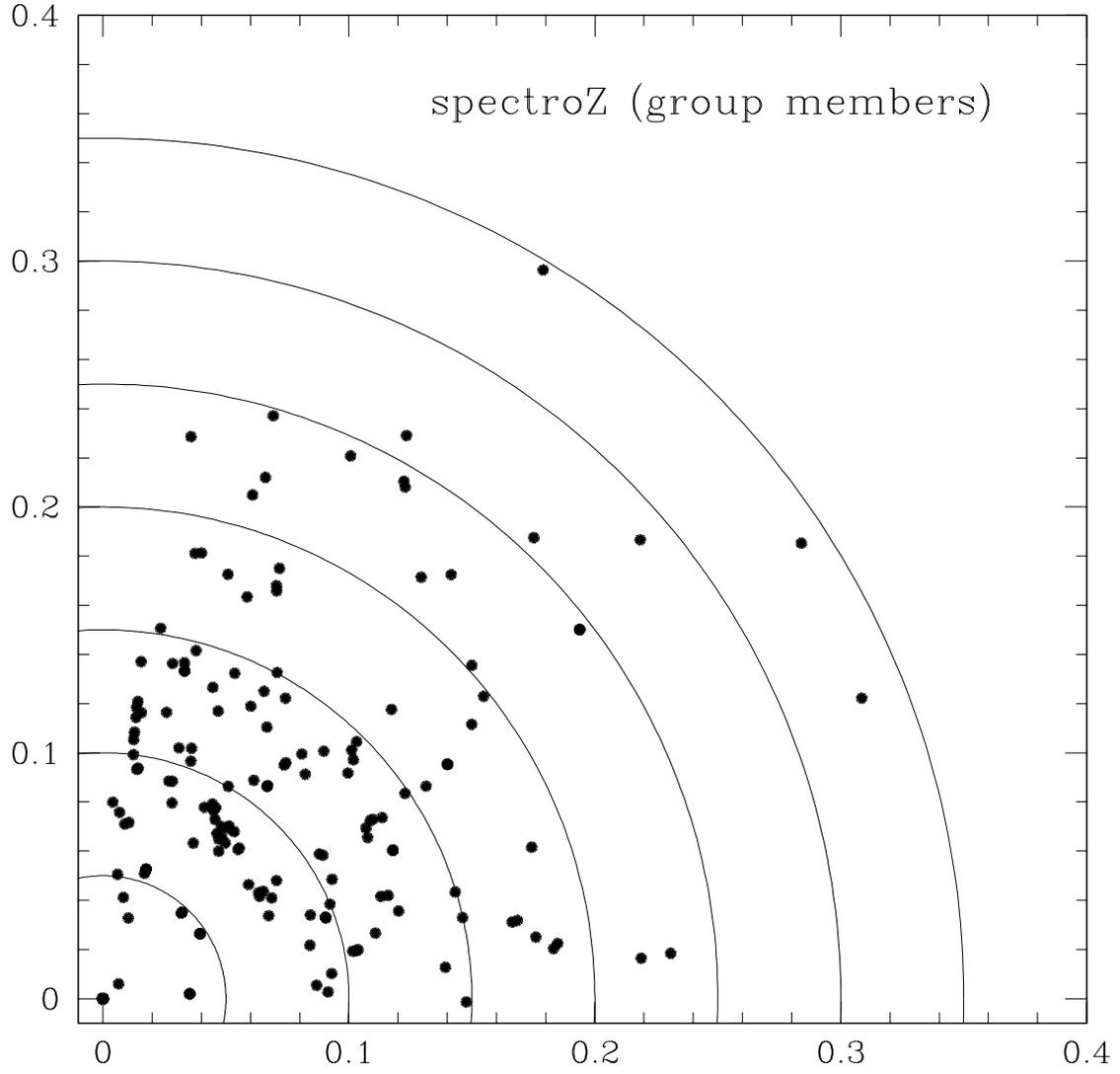}
\caption{The space distribution of individual 158 CG galaxies with 
spectroscopic redshift (redshift/RA wedge plot). \label{fig5}}
\end{figure*}

As noted at the beginning of this section, only 158 of the 744 CG
member galaxies have spectroscopic redshifts either from the SDSS or
from the literature (see Fig.~\ref{fig5}).  Our magnitude cut for CG
galaxies is $r^*_{\rm model} \leq 21.0$ (\S~\ref{cat_construct}),
whereas the magnitude limit for the main galaxy sample of the SDSS
spectroscopic program is $r^*_{\rm petro} \leq 17.7$ [Table~29 of
\citet{Stoughton02}].  Further complicating matters  is an SDSS single
plate target proximity limit  of 55~arcsec  to avoid fiber  collisions
\citep{Blanton03a}, which in general would allow only a single galaxy 
from a compact group to be targeted on a spectroscopic plate.
Therefore, it is not surprising that so few of our CG galaxies have
spectroscopically determined redshifts.  Nonetheless, these 158
galaxies are spread fairly uniformly over the SDSS CG sample; so, a
full 75\% of our CGs (131 out of 175) contain at least one group
member with a spectroscopically determined redshift.

We were, however, able to obtain a photometric redshift for every SDSS
CG (see Table~\ref{tab_cg}).   We did this  by taking the  photometric
redshift of the brightest group  member as the photometric redshift of
the CG as a whole.  These brightest group member photometric redshifts
were   determined  by   a  variation  on   the   polynomial  method of
\citet{Connolly95}.  Instead of   using polynomials,  however,  Pad\'e
approximants \citep{Gershenfeld99}  were used.    Pad\'e approximants,
which have   seen  applications  in statistical  mechanics,   critical
phenomena, and circuit design \citep{Baker96}, are basically ratios of
polynomials.  As a training sample, $\approx$ 500 main sample galaxies
\citep{Strauss02} and $\approx$    500 Luminous Red   Galaxies  (LRGs)
\citep{Eisenstein01}  were  taken from  the SDSS spectroscopic sample;
the low-redshift  training   set   sample was supplemented    with  an
additional  $\approx$ 500 galaxies  from a special spectroscopic plate
which attempts to sample  fully from the bright  end of the luminosity
function at $z<0.15$  \citep{Lin03}.  Details on   this method can  be
found elsewhere \citep{Annis03}.

In calculating the photometric  redshifts of the  CGs, we ignored  the
non-first-ranked group  members since using them  tended to  result in
systematic errors of $\Delta z = 0.1$ or more.  This  is likely due to
the fact that 85\% of the  non-first-ranked galaxies in our sample are
fainter than $r^* = 17.7$, where  the training set for our photometric
redshift relation  is   dominated by LRGs,  a  type  of   galaxy which
probably does  not describe these fainter  CG galaxies very  well as a
whole.  In contrast, only 16\%  of brightest group members are fainter
than $r^* = 17.7$ ---  and none fainter  than $r^* = 18$; furthermore,
since LRGs have luminosities and spectral energy distributions similar
to brightest cluster members, the  LRG training set is probably better
suited  to our  brightest group members    than for the fainter  group
members.

\begin{figure*}[ht]
\centering
\vspace{5.0cm}
\includegraphics[angle=-90,scale=0.6]{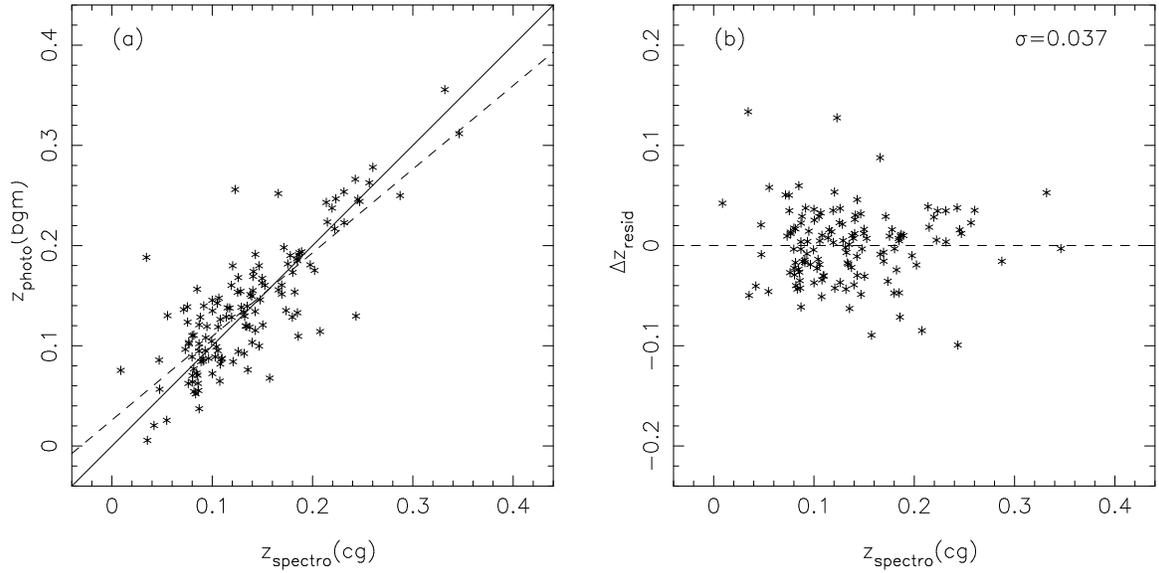}
\vspace{2.0cm}
\caption{(a) The photometric redshift of the brightest group member,
$z_{\rm photo}^{\rm bgm}$,  vs.    the spectroscopic redshift  of  the
group,  $z_{\rm spectro}^{\rm cg}$, for the  set of 131 SDSS CGs which
contain at least one  group member with a spectroscopically determined
redshift.  The  solid  line depicts the  relation  $z_{\rm photo}^{\rm
bgm}   = z_{\rm spectro}^{\rm cg}$;     the dashed line indicates  the
unweighted least squares fit, $z_{\rm photo}^{\rm bgm} = 0.026 + 0.834
\times  z_{\rm spectro}^{\rm  cg}$.  (b)  The  residuals of the   least
squares fit in (a).\label{fig:xxx1}}
\end{figure*}

\begin{figure*}[ht]
\centering
\vspace{5.0cm}
\includegraphics[angle=-90,scale=0.6]{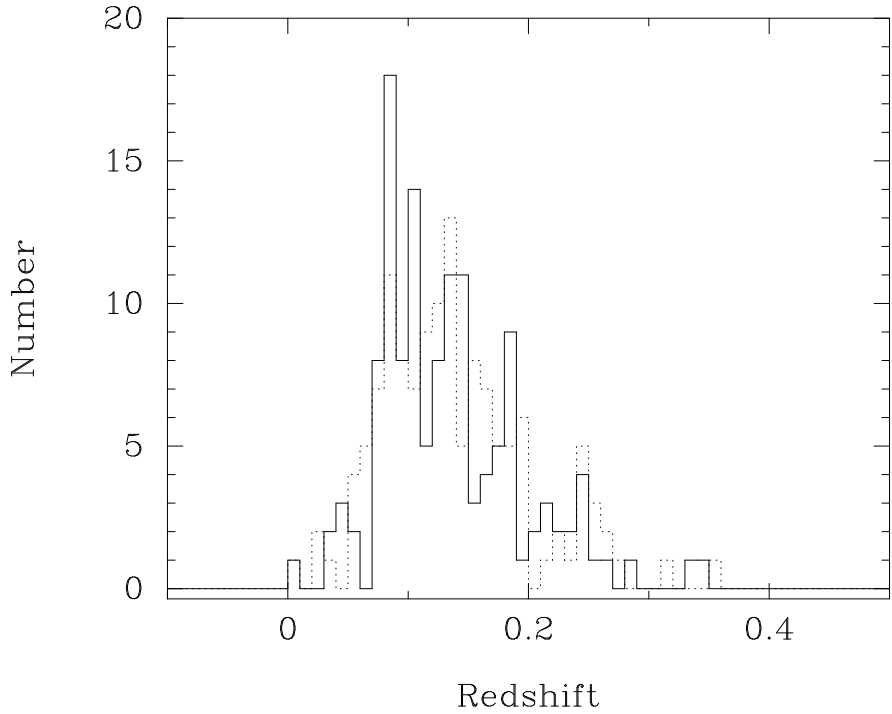}
\vspace{2.0cm}
\caption{The distribution of redshifts for those 131 SDSS CGs with a
spectroscopically   determined redshift.  The     solid line   is  the
histogram of  spectroscopic redshifts, $z_{\rm spectro}^{\rm cg}$, for
this  sample of 131 SDSS  CGs and  has  a mean of  $0.135\pm0.005$ and a
median of 0.126.  The dotted line is the histogram of photometric redshifts,
$z_{\rm photo}^{\rm bgm}$, for this sample of 131 SDSS CGs and has a mean 
of $0.138\pm0.005$ and a median of 0.130.\label{fig:xxx2}}
\end{figure*}

How  well does a brightest group  member photometric redshift, $z_{\rm
photo}^{\rm bgm}$, track  the group's spectroscopic redshift,  $z_{\rm
spectro}^{\rm cg}$?   In   Figure~\ref{fig:xxx1}a,  we  plot   these   two
parameters against each other for those 131 SDSS  CGs which contain at
least  one member with  a spectroscopic redshift.  The relationship is
quite good.  The  rms of the residual  of the unweighted least squares
fit  of brightest group  member   photometric redshift  to the   group
spectroscopic redshift is $\sigma_{z} = 0.037$ (Fig.~\ref{fig:xxx1}b),
which means  that the typical $1\sigma$ rms  error  in the photometric
redshift of an individual  group is about  0.037.  Further, if we plot
the  histograms      of  $z_{\rm  photo}^{\rm   bgm}$      and $z_{\rm
spectro}^{\rm cg}$,  we find  that,   overall, the two  distributions look
quite similar (Fig.~\ref{fig:xxx2}).

\begin{figure*}[ht]
\centering
\vspace{5.0cm}
\includegraphics[angle=-90,scale=0.6]{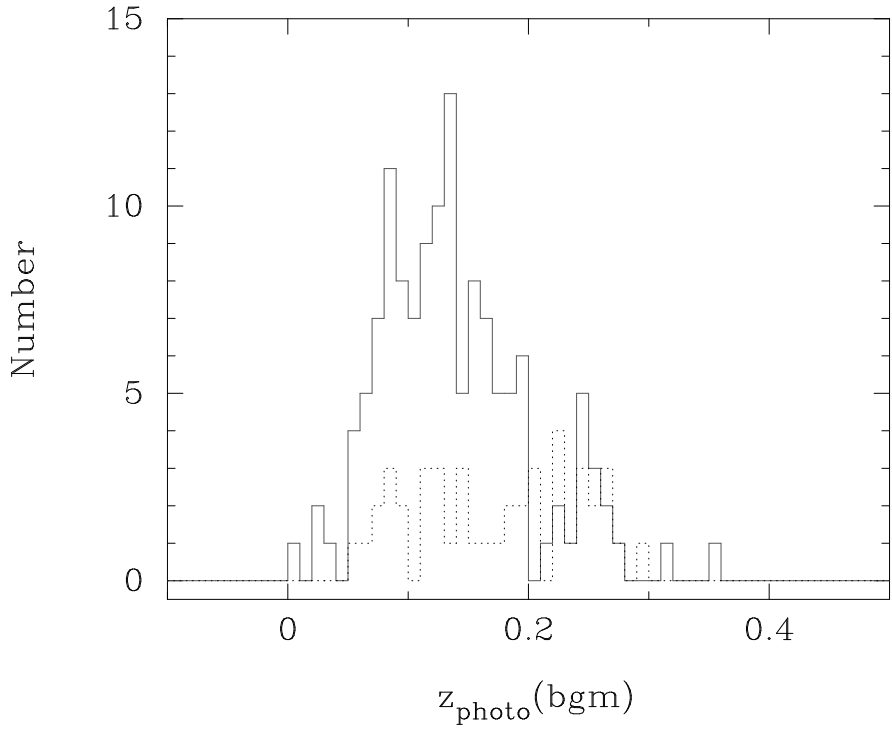}
\vspace{2.0cm}
\caption{The solid line is the histogram of photometric redshifts for 
the  sample of  131 SDSS CGs   that  have spectroscopically determined
redshifts;  it  is    the same   as    the  dotted line    plotted  in
Fig.~\ref{fig:xxx2}.  The dotted line is  the histogram of photometric
redshifts for the  44 SDSS CGs  that do  not have  a spectroscopically
determined redshifts; this  distribution has a mean of $0.174\pm0.010$
and  a median of  0.178.  The full  sample of 175 SDSS  CGs has a mean
photometric redshift  of   $0.147\pm0.005$ and a   median  photometric
redshift of 0.136.\label{fig:xxx3}}
\end{figure*}

Based  on  the  131  SDSS CGs  with  spectroscopic   redshifts, we can
estimate that the median redshift of  the full sample  of 175 SDSS CGs
to be  $<$$z_{\rm spectro}$$>$$_{\rm   med}  = 0.126$.      This estimate,
however, may be biased:  25\% of the SDSS  CGS in  the full sample  of
have no spectroscopic redshifts and  this subset may be systematically
more  (or  less)   distant than  the   rest of   the   SDSS  CGs.   In
Figure~\ref{fig:xxx3}   we  plot   the  distribution   of  photometric
redshifts for the 131 SDSS  CGs  with spectroscopic redshifts and  the
distribution  of photometric redshifts for   the  44 SDSS CGs  without
spectroscopic  redshifts.  Indeed,  the  set  of  44 SDSS CGs  without
spectroscopic redshifts appears to  be, on average, more distant  than
the set of 131 galaxies with spectroscopic redshifts.   Even so, it is
a smaller set, and, if we combine both sets of CGs, we find the median
photometric redshift  for the  full   sample of 175   SDSS  CGs to  be
$<$$z_{\rm photo}$$>$$_{\rm med} = 0.136$.   This value is only marginally
larger    than  our estimate    for  $<$$z_{\rm spectro}$$>$$_{\rm  med}$.
Therefore,  unless otherwise  noted, we   will  use the  spectroscopic
redshifts throughout the remainder of the paper.

\subsection{The Local Environment} 

Following \citet{Palumbo95}'s  and \citet{Iovino02}'s studies  of  the
local  environments surrounding  HCGs and SCGs,  we  have undertaken a
similar study of the local environments surrounding SDSS CGs.

For each SDSS CG  we searched a  circular region $5\times$ the group's
radius $\theta_G$ in  the  SDSS  EDR  data base  for  galaxies in  the
interval  $r^*_{\rm bgm} < r^*  < r^*_{\rm fgm}  + 1$, where $r^*_{\rm
bgm}$  and $r^*_{\rm fgm}$ are the   SDSS $r$-band model magnitudes of
the brightest and the faintest  group members, respectively, for  that
particular group, as   listed in  Table~\ref{tab_cgmem}.  The  surface
density of the SDSS CG  ($\rho^{\rm SDSSCG}$) was obtained by dividing
the number    of    these  galaxies   within    1  group   radius   by
$0.25\pi\theta^2_G$.  The surface  density  of galaxies in  that  SDSS
CG's local environment $\rho^{\rm env}$ was  calclated by dividing the
number of these  galaxies  in an  annulus $3 -  10\times$  the group's
radius by the area of this annulus.

\begin{figure*}[ht]
\centering
\includegraphics[angle=0,scale=0.6]{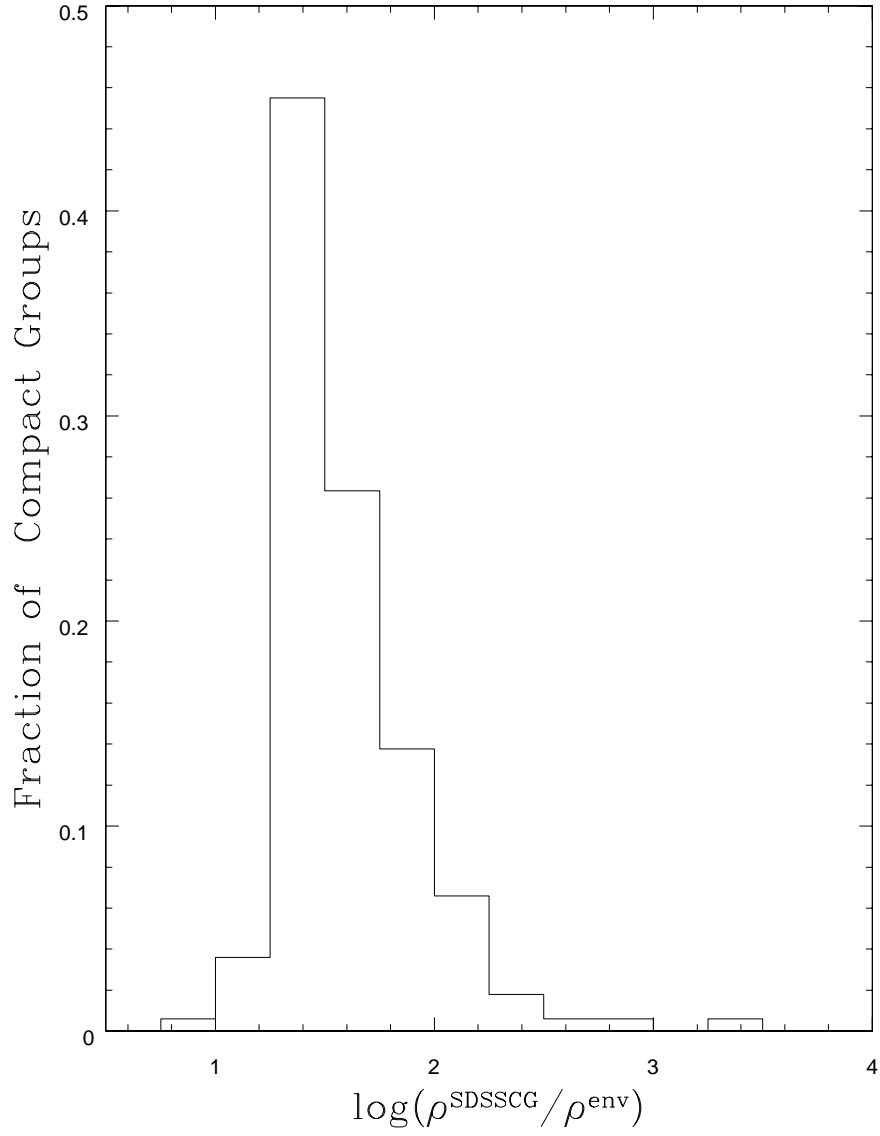}
\caption{The distribution  of the surface  density  ratio log($\rho^{\rm SDSSCG}/\rho^{\rm env}$)
for SDSS CGs.\label{fig:pho1}}
\end{figure*}

The distribution of   the ratio $\rho^{\rm SDSSCG}/\rho^{\rm env}$  is
shown in Figure~\ref{fig:pho1}.   Note  that the surface  densities of
SDSS CGs are factors of $5  - 3000$ greater  than those of their local
environments  (on average,  they are  about  a factor of 40  greater),
similar    to  what  \citet{Palumbo95}   found    for   HCGs and  what
\citet{Iovino02} found for SCGs.

\begin{figure*}[ht]
\centering
\includegraphics[angle=0,scale=0.6]{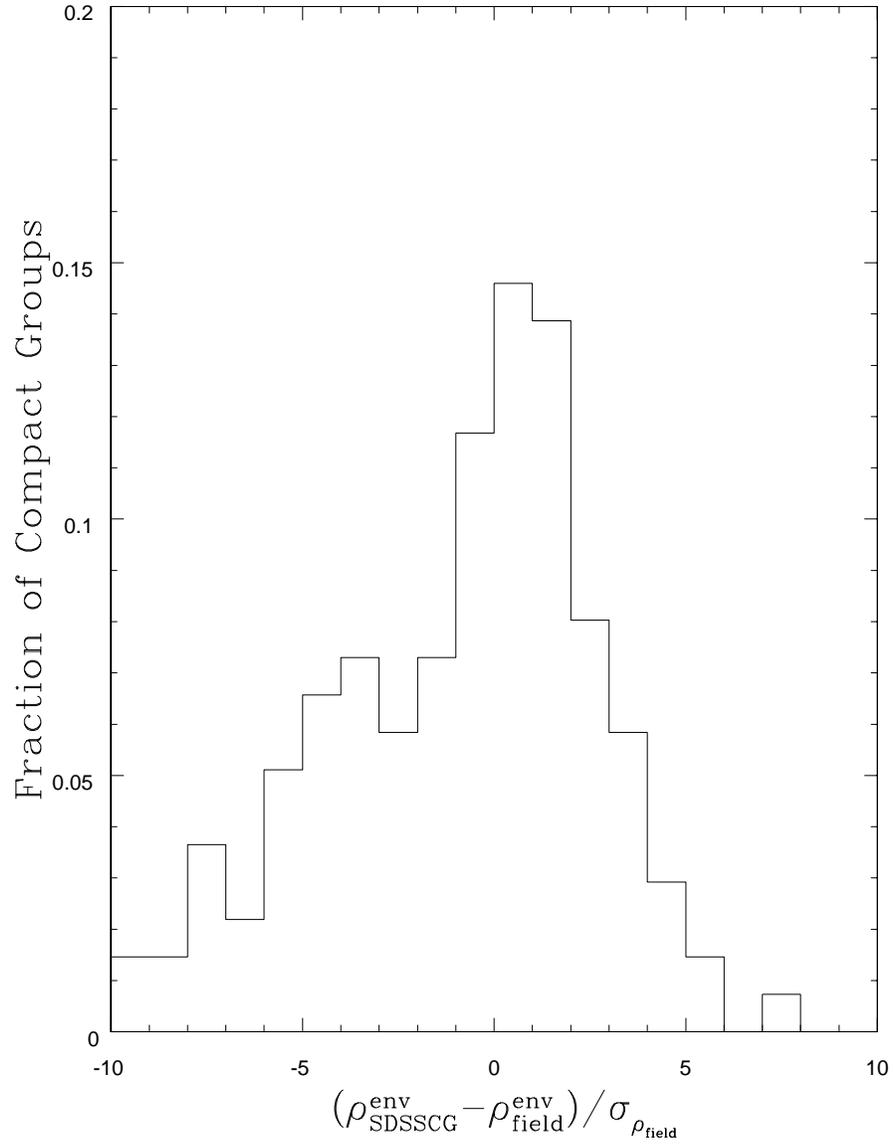}
\caption{The distribution  of the local SDSS CGs environments relative to that of the 
field galaxies.\label{fig:pho2}}
\end{figure*}

To test  how the local  environment of SDSS CGs  compares with that of
the field, we  considered the local environments  of galaxies from the
SDSS isolated galaxy catalog  \citet{Allam03}.  For  each SDSS CG,  we
selected isolated galaxies of comparable  brightness to the  brightest
group member ($r^*_{\rm bgm} -0.5 < r^{*} < r^*_{\rm  bgm} + 0.5$) and
observed under  comparable seeing conditions.   We then calculated the
surface  density  of galaxies surrounding  of  each  of these isolated
galaxies in an annulus $3 - 10\times$ the group's radius, just like we
did to calculate $\rho^{\rm  env}$ for the   SDSS CG in question.   We
then calculated  the mean value of the  field $\rho^{\rm env}$ and its
scatter  $\sigma_{\rho_{\rm field}}$.  Figure \ref{fig:pho2} shows the
histogram of the quantity $(\rho^{\rm env}_{\rm SDSSCG} -
\rho^{\rm env}_{\rm field})/\sigma_{\rho_{\rm field}}$.
We find  that the environments of  SDSSCGs are, on average, similar to
the  environments of field  galaxies,  although there is  considerable
scatter.    This    result  is    comparable   to   those    found  by
\citet{Palumbo95} for HCGs and by \citet{Iovino02} for SCGs.


\section{Comparison with other CGs} 
\label{comparison_other}

It  is instructive to compare the  properties of  SDSSCGs with those of CGs
from   other catalogs.   In  Table~\ref{tab:Cgstat}  we  summarize the
properties of the current  SDSSCG catalog and  those  of six other  CG
catalogs  ---   the initial  DPOSS  Compact    Group (PCG)  catalog by
\citet{Iovino03},   the  Southern   Compact Group    (SCG) catalog  by
\citet{Iovino02},   the  UZC   Compact   Group  (UZC-CG)    catalog by
\citet{FocardiKelm02}, the Las Campanas  Compact Group  (LCCG) catalog
by  \citet{AllamTucker00}, the Redshift  Survey  Compact Group  (RSCG)
catalog  by \citet{Barton96},  and  the  Hickson  Compact  Group (HCG)
catalog by \citet{Hickson82,Hickson93}.

Of  these seven CG catalogs, four  are based upon galaxy sky positions
and  photometry alone (the SDSSCG,  PCG,  SCG, and  HCG catalogs), and
three also make use of  galaxy redshift information (the UZC-CG, LCCG,
and RSCG catalogs).

\begin{figure*}
\centering
\includegraphics[angle=-90,scale=0.7]{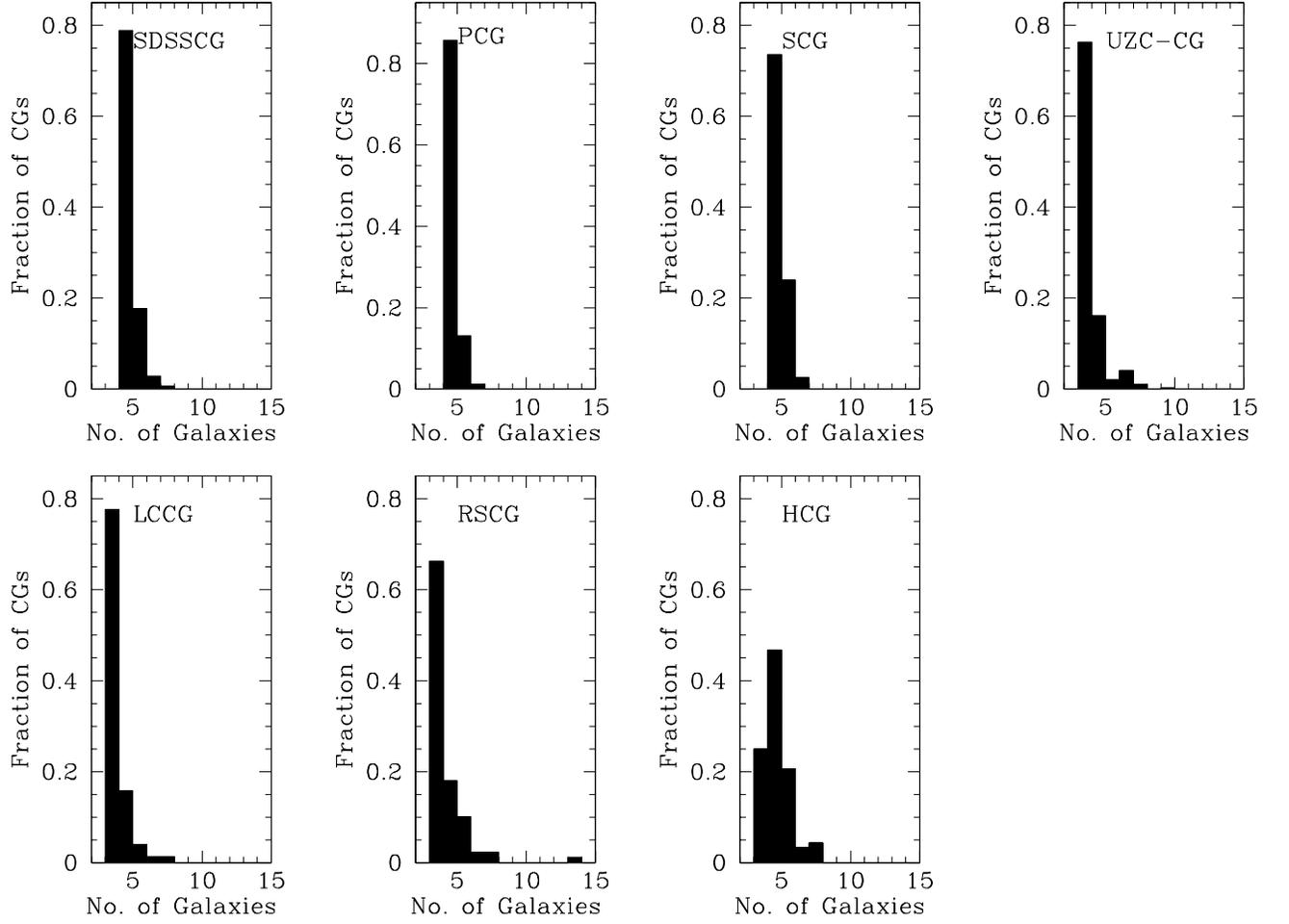}
\caption{The distribution of group memberships, $N$, for the SDSSCG, 
PCG, SCG, UZC-CG, LCCG, RSCG, and HCG catalogs.
\label{statcgsN}}
\end{figure*}

Looking at this table and its companion figures (Figs. \ref{statcgsN},
\ref{statcgsZ}, \& \ref{statcgsThetaG}), it  is apparent  that SDSSCGs
are quite similar to most other CG catalogs  in terms of their typical
group membership ($N$)  and linear group  diameter ($D\times\theta_G$;
Fig.~\ref{statcgsThetaG}).  With the possible exception of the PCG catalog, 
however, the SDSSCG catalog is on average 
much  deeper than any  of the other CG catalogs.   (Although we do not
have  redshift information  for the SCG   catalog, the mean and median
angular diameters of its groups, coupled with a galaxy magnitude limit
of $b_j  = 15.0$, indicate that  it has a depth  comparable to that of
the  RSCG catalog.)  With such a  mean depth, the  SDSSCG catalog is a
first step  toward the study  of the  evolutionary properties of CGs.
Furthermore, at these depths, the probability of finding CGs acting as
strong lenses becomes  noticeably higher \citep{Hickson97};  perhaps a
CG from  a future SDSSCG catalog  will be the  first  to be discovered
with giant arcs.

\begin{figure*}
\centering
\includegraphics[angle=0,scale=0.8]{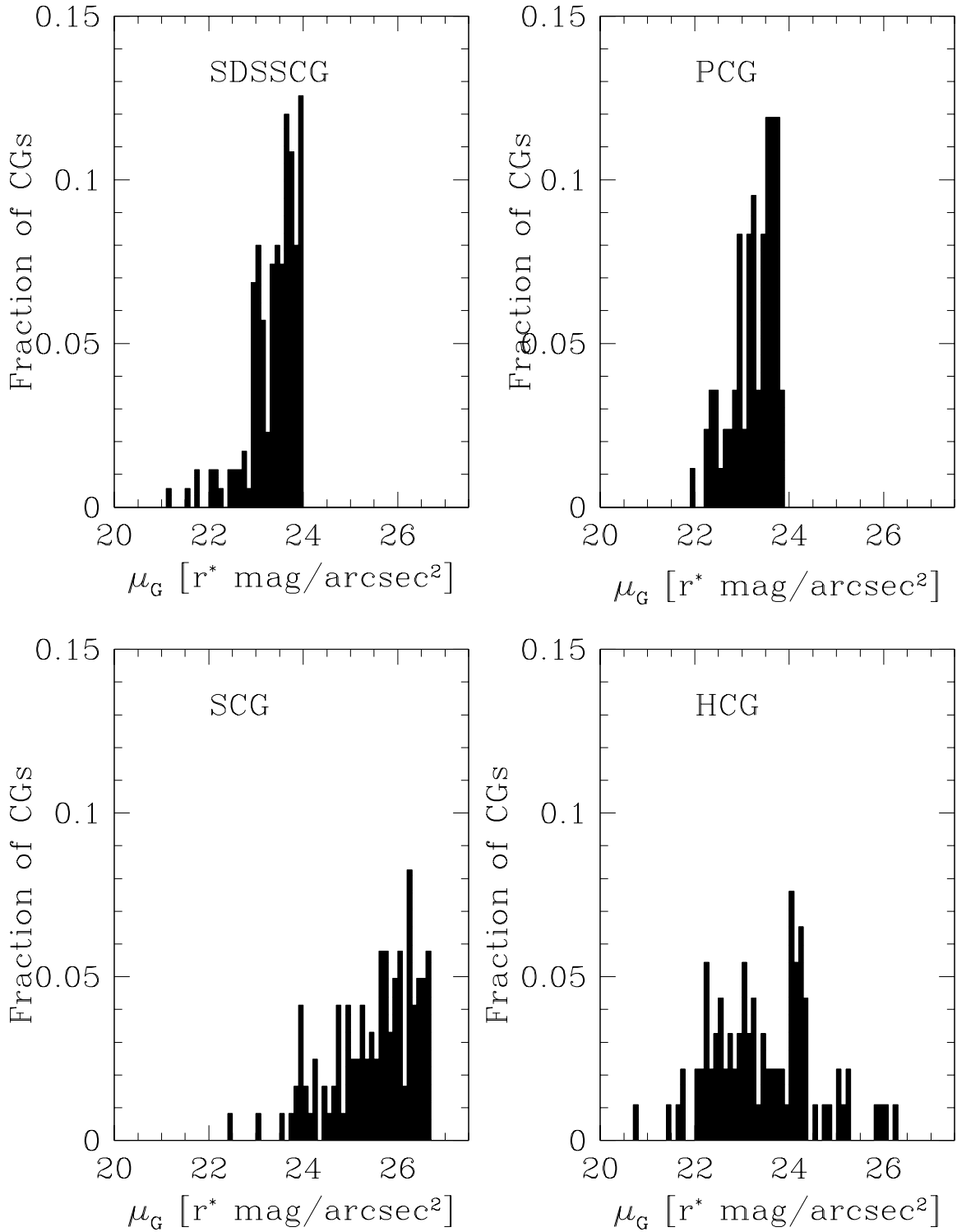}
\caption{The distribution of group surface brightness, $\mu_{G}$ ~mag~arcsec$^{-2}$, 
in the SDSS $r^{*}$--band for the SDSSCG, PCG, SCG, and HCG  catalogs.
\label{statcgsmugall}}
\end{figure*}

\begin{figure*}
\centering
\includegraphics[angle=0,scale=0.8]{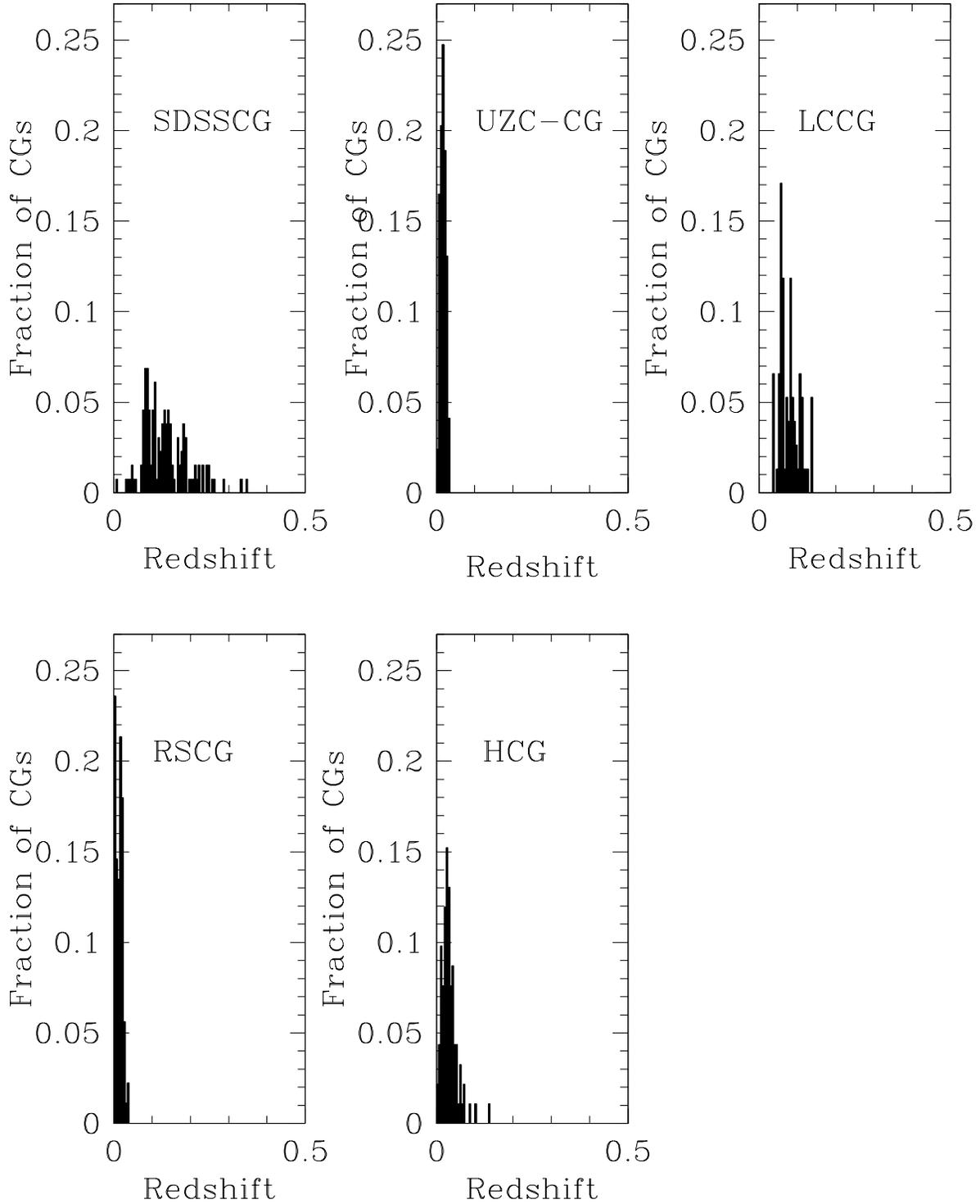}
\caption{The distribution of group redshifts, $z$, for the SDSSCG, UZC-CG,
LCCG,  RSCG, and HCG  catalogs.  Note  that  the SDSSCG  values  make use  of
only those 131 CGs with known spectroscopic redshifts. 
\label{statcgsZ}}
\end{figure*}

\begin{figure*}
\centering
\includegraphics[angle=0,scale=0.8]{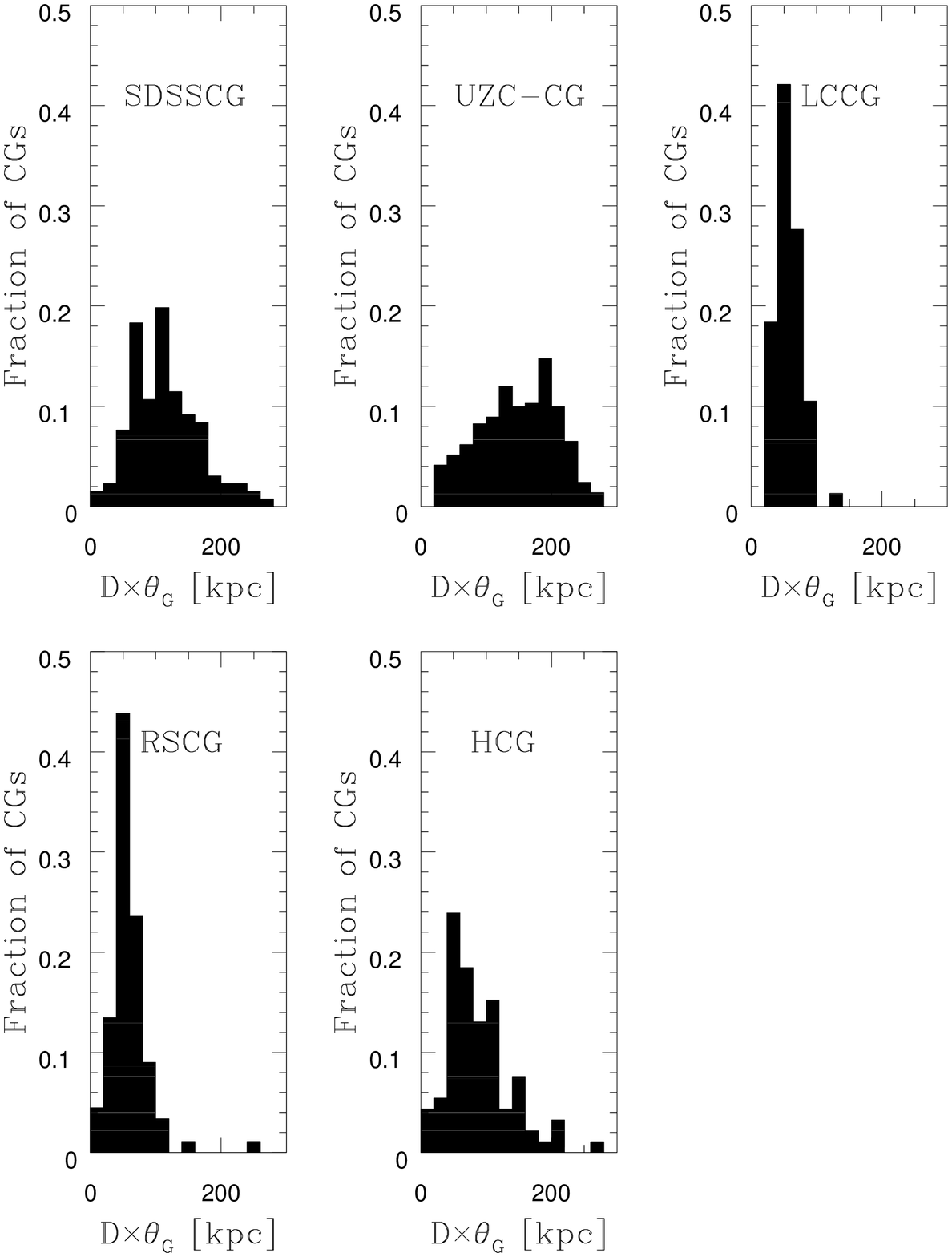}
\caption{The distribution of linear group diameters, $D\times\theta_G$ 
($H_0=100$~km~s$^{-1}$~Mpc$^{-1}$), for   the SDSSCG,  UZC-CG,   LCCG,
RSCG, and HCG  catalogs.  Note  that  the SDSSCG  values  make use  of
only those 131 CGs with known spectroscopic redshifts.
\label{statcgsThetaG}}
\end{figure*}

It is  also  curious  to see that,   although  we  used  a compactness
criterion  to detect SDSSCGs  that was substantially  tighter than the
original  \citet{Hickson82}   value,  the mean   surface  brightnesses
($\mu_G$) in $r^*$ for SDSSCGs and HCGs  are nearly identical.  Recall
that the HCG catalog  was extracted from the  first Palomar Sky Survey
(POSS-I)   via   a  search-by-eye.   \citet{Hickson82}  expected,  and
\citet{Prandoni94} confirmed,  that  the  HCG catalog  was  incomplete
approaching the limits of the Hickson criteria.  Here, we appear to be
seeing an independent confirmation of the HCG catalog's incompleteness
for ``less compact'' CGs as clearly seen in Fig \ref{statcgsmugall}.

Finally, although a detailed calculation of the space  density of CGs is beyond
the    scope  of  this  paper,  we  can    perform a rough,
``back-of-the-envelope'' estimate  by assuming that  a given CG sample
is complete out to its median redshift and calculating
\begin{equation}
n_{\rm cg} = \frac{0.5 N}{(A/41253{\mbox{ sq deg}}) \frac{4}{3} \pi D_{\rm c}^3} 
\label{eqn:ncg}, 
\end{equation}
where $n_{\rm cg}$ is the estimated CG space density, $N$ is the total
number  of CG's in  the  sample, $A$ is the   sky area covered  by the
sample in square degrees, and $D_{\rm c}$  is the co-moving distance at
the median redshift of the survey.  We use  a Euclidean measure of the
volume   since   we will assume  a    zero-curvature Universe in which
$\Omega_{\rm M} = 0.3$ and $\Omega_{\Lambda}=0.7$ (see, for example,
\citealt{Hogg00}).

In particular, we wish to compare  the space density  of SDSS CGs with
that of  the PCG catalog,  which is presently the  only other deep ($z
\ga 0.1$), wide area survey for CGs, and with that of the HCG catalog,
which remains the benchmark for all CG catalogs.

For the SDSS CGs, we take $N=175$,  $A=253$~sq~deg, and $z_{\rm med} =
0.126$ ($D_{\rm   c} = 368h^{-1}$~Mpc),   yielding an  estimated space
density  of  $1.1   \times  10^{-4}h^3$~Mpc$^{-3}$.   Using  the above
equation, we can  also estimate the  space density of the PCG catalog.
For the  PCGs, we take  $N=84$, $A=2000$~sq~deg, and $z_{\rm med}=0.1$
($D_{\rm  c}= 280h^{-1}$~Mpc), yielding  an estimated space density of
$9.4 \times  10^{-6}h^3$~Mpc$^{-3}$.   Clearly,  the space  density of
SDSS CGs is about a factor of ten greater than that for PCGs.  The PCG
catalog, however, has  a much  tighter  mean group surface  brightness
constraint.  If we consider just the 14 SDSS  CGs that satisfy the PCG
surface brightness criterion, we find a space density for ``PCG-like''
SDSS CGs of $9.0 \times 10^{-6}h^3$~Mpc$^{-3}$ --- almost identical to
the  value obtained  for the  original  PCG sample.   This is  a  good
consistency check between these two modern CG catalogs.

Comparing the space density of CGs with that  of HCGs is a bit harder.
Since the HCG catalog is known to  be incomplete, we cannot merely use
equation \ref{eqn:ncg}.  Using simulations, \citet{MendesDeOliveira91}
estimated  that     the      space     density  for     HCGs        is
$3.9\times10^{-5}h^3$~Mpc$^{-3}$, or about one-third the space density
of SDSS  CGs.  Worse  yet,  SDSS CGs  have  a more  stringent  surface
brightness criterion.  We  find that 63\% (58  out of 92) of HCGs meet
the SDSS CG surface brightness constraint.  Thus, the space density of
SDSS CG-like HCGs should be about $2.5\times10^{-5}h^3$~Mpc$^{-3}$, or
only  about one-fifth  to one-fourth the   space density of SDSS  CGs.

Four possibilities  come readily to  mind to explain this discrepancy.
(1) The  HCG catalog may   still be substantially incomplete  even for
surface brightnesses of $\mu_G(r^*)  < 24.0$.  If  so, a more complete
sample of  HCG-like  groups would have  a  higher  space density.  (2)
\citet{MendesDeOliveira91}'s estimate of the space  density of HCGs is
itself  uncertain and is very  sensitive  to the assumptions regarding
the completeness of the  HCG  sample \citep{Hernquist95}.   Therefore,
the discrepancy may not be very statistically significant.  (3) Cosmic
variance may  play an effect.  Out  to their median redshifts, the HCG
and the  SDSS CG volumes   are both roughly  equivalent to  boxes only
100$h^{-1}$~Mpc on a side.  (4) Although  much care was taken to avoid
contamination by chance projections (\S~\ref{cat_construct}), the SDSS
CG catalog itself  probably   contains some small fraction   of spurious 
groups, thus  artificially  increasing our estimated space  density of
SDSS  CGs.    We  suspect    that  some combination  of   these   four
possibilities will eventually explain the present discrepancy.


\section{Morphology-Environment Effects in SDSS CGs}
\label{morph-env}

To search for environmental     effects  on the morphologies  of    CG
galaxies, we  have used  the sub-sample of  158 SDSSCG  galaxies which
have  known spectroscopic redshifts.  In order  to compare this sample
fairly, we have  constructed a field  sample by selecting for  each of
these 158 SDSSCG galaxies the ten galaxies  with the nearest redshifts
to it  from the SDSS  EDR \citep{Eisenstein01,Strauss02}.  Since these
ten galaxies  will tend to be  randomly spread in  RA and DEC over the
SDSS  EDR  survey area, they will  reasonably   sample the  field with
little  contamination from clusters.  (Recall that  only about 10\% of
galaxies lie within  rich clusters).  We have  thus collected a  field
sample of 1580 galaxies which has  a redshift distribution essentially
identical  to that of the original  158 SDSSCG  galaxies.  That we met
this goal is attested to by the fact that the means and medians of the
two         redshift         distributions         are       identical
(Table~\ref{tab:MeanMedianCompare}).  Furthermore,   a one-dimensional
Kolmogorov-Smirnov (KS) test \citep{Press92}  indicates that these two
redshift  distributions are  drawn from  the  same parent distribution
with an extremely high probability (Table~\ref{tab:1DKS}).

A galaxy's color is a function of current star formation rate, and, as
such, its color  can be used  as  a rough indication of  morphological
type.  Red galaxies   tend to be  Ellipticals and  S0's; blue galaxies
tend to be Spirals and Irregulars.  The SDSS data set contains not one
but five filters' worth of photometry, which can  define a set of four
non-redundant  colors:    $(u^*-g^*)$,   $(g^*-r^*)$, $(r^*-i^*)$, and
$(i^*-z^*)$.   We  will  use this plethora    of  color information to
compare our two samples.

For our   comparisons, we construct   Color-Color and  Color-Magnitude
Diagrams for these  two samples.   To  do this, though, we  must first
correct the  galaxy magnitudes and  colors for interstellar absorption
and for cosmological effects.  To correct for interstellar absorption,
the following reddening corrections
\citep{Stoughton02} were subtracted from the galaxy magnitudes:
\begin{eqnarray}
A(u^*) & = & 5.155 \times E(\bv) \nonumber \\
A(g^*) & = & 3.793 \times E(\bv) \nonumber \\
A(r^*) & = & 2.751 \times E(\bv) \\
A(i^*) & = & 2.086 \times E(\bv) \nonumber \\
A(z^*) & = & 1.479 \times E(\bv) \nonumber
\end{eqnarray}
where  the    values   for    $E(\bv)$  were    obtained     from  the
\citet{Schlegel98} reddening maps.  (Note   that we correct only   for
interstellar absorption due to the Milky Way Galaxy and do not attempt
to correct for internal interstellar   absorption associated with  the
individual galaxies themselves.)

We also apply  a k-correction to  convert  the de-reddened colors  and
magnitudes  to their rest-frame   (i.e., redshift  $z=0$) equivalents.
Our  k-corrections  were estimated  using  the publicly available {\tt
kcorrect} (v1.10) package of   \citet{Blanton03b}.   This code  fits  a
galaxy's (redshifted) broad-band   magnitudes to a suite of   template
galaxy   spectral energy distributions in   order  to reconstruct  the
galaxy's broad-band  magnitudes   at  another redshift (e.g.,    at  a
redshift of $z=0$).

Finally, to   calculate absolute $r^*$ magnitudes,  we  must  assume a
cosmological  model.  We take the   current standard ---  a flat model
with $\Omega_{\rm   M} = 0.3$, $\Omega_{\Lambda}  =  0.7$,  and $H_0 =
100h$~km~s$^{-1}$~Mpc$^{-1}$ --- using the analytical relation by
\citet{Pen99} to calculate luminosity distance $d_{\rm L}$.

Figure~\ref{gr_vs_z}        plots   the de-reddened  and   k-corrected
$M_{g^*}-M_{r^*}$  color  against spectroscopic  redshift for  our two
samples.   The  E/S0  ridgeline ---  the  locus of  Ellipticals and S0
galaxies at  $M_{g^*}-M_{r^*} \approx 0.8$ ---  is nearly flat for the
redshift range plotted, indicating  that the k-corrections did a  good
job.  The  slight positive slope  of  the ridgeline between  $z=0$ and
$z=0.3$ may  be the result of a  slight  color evolution in early-type
galaxies over this redshift range.

Figures~\ref{mag_vs_color}  and   \ref{color_vs_color}     display the
Color-Magnitude and the  Color-Color  Diagrams  for the  two  samples,
respectively.   Visually,  it is hard   to distinguish much difference
between the  SDSSCG and field galaxy samples  in these plots, with the
possible exception that the SDSSCG galaxies appear on average slightly
redder  than   the  field    galaxies  in  $M_{u^*}-M_{g^*}$   and  in
$M_{g^*}-M_{r^*}$.

If, however,  we look at the statistics  of the absolute magnitude and
rest-frame color  distributions,  differences between the  two samples
become more apparent.  If  we look at the means  and medians  of these
distributions for the two samples (Table~\ref{tab:MeanMedianCompare}),
it is clear that   the SDSSCGs are  on  average redder than  the field
galaxies in $M_{u^*}-M_{g^*}$ and  in  $M_{g^*}-M_{r^*}$ at about  the
$2\sigma$ level.  Furthermore, there is some evidence --- at about the
$1\sigma$  level ---  that SDSSCGs  are on average  less luminous than
field galaxies (in the $r^*$-band).

We have  also run two sets  of KS tests  \citep{Press92} for these two
samples: (1) a set of one-dimensional KS tests on the distributions of
$z$,    $M_{r^*}$,          $M_{u^*}-M_{g^*}$,      $M_{g^*}-M_{r^*}$,
$M_{r^*}-M_{i^*}$,  and $M_{i^*}-M_{z^*}$ (Table~\ref{tab:1DKS});  and
(2)  a set   of  two-dimensional  KS tests    on the plots   shown  in
Figures~\ref{gr_vs_z} -- \ref{color_vs_color} (Table~\ref{tab:2DKS}).

These two sets of KS tests provide strong evidence that the rest-frame
colors of  CG galaxies do indeed  differ from  those of field galaxies
---  at   least for   $M_{u^*}-M_{g^*}$,  $M_{g^*}-M_{r^*}$,  and even
$M_{r^*}-M_{i^*}$.   The  distribution of $M_{i^*}-M_{z^*}$ rest-frame
colors, however, does  not appear to  differ significantly between the
two galaxy samples.  This is not surprising.  The intrinsic rest-frame
$M_{i^*}-M_{z^*}$ color of an Elliptical is only about 0.25~mag redder
than that of an  Irregular; in comparison, the $M_{u^*}-M_{g^*}$ color
of an Elliptical is about 1.3~mag redder than that of an Irregular
\citep{Fukugita95}.  Thus, the $M_{i^*}-M_{z^*}$ colors are not a
very sensitive measure of galaxy morphology.

From the evidence  of their $M_{u^*}-M_{g^*}$,  $M_{g^*}-M_{r^*}$, and
$M_{r^*}-M_{i^*}$ rest-frame colors,  we conclude that SDSSCGs contain
a relatively  higher fraction of   Elliptical galaxies than  does  the
field.

Why   does the relative  fraction of  Elliptical  galaxies in SDSS CGs
appear to be  larger  than in  the field?   A  likely culprit  is  the
cumulative effect of  interactions and mergers   over the course  of a
typical CG lifetime.  N-body simulations like those pioneered by
\citet{Toomre77} indicate that the end-product of merging Spirals 
can be an Elliptical galaxy.  

There is evidence for interactions and/or mergers  in SDSS CGs.  While
culling the original set of 232 candidate SDSS CGs to the final set of
175   (\S~\ref{cat_construct}),  the   member galaxies  were  visually
inspected for   tidal tails, bridges,   and   obvious other signs   of
interaction activity.  Those SDSS CGs showing evidence of interactions
and/or mergers are indicated in the ``Comments'' column (Column~12) of
Table~\ref{tab_cg}.  Evidence of   some sort of  interaction or  tidal
tail is  seen in 55 SDSS  CGs (31\% of the final  catalog  of 175 SDSS
CGs); full-blown merger events appear  to be occurring  in 26 SDSS CGs
(14\%).  A  luminous halo enveloping the all  or part of the group ---
perhaps the remnant of a completely disrupted  galaxy --- appears in 4
of the SDSS CGs (2\%).

\cite{Zepf1993} estimated that roughly 7\% of the galaxies in HCGs are
in  the process of   merging.  His conclusion    was based on  roughly
consistent frequencies of (a) optical signatures  of merging, (b) warm
far-infrared colors, and  (c) sinusoidal rotation curves.  Our initial
results, described  above, indicate comparable  or even greater levels
of merger activity in the SDSS CG sample.


\section{Conclusions}
\label{conclusions}

We have presented a new catalog of CGs, one extracted via an objective
algorithm from   153~sq~deg  of Runs  752   and 756  in  the  SDSS EDR
\citep{Stoughton02}.    The  algorithm,  using  a    modified form  of
Hickson's  \citeyearpar{Hickson82} original criteria, detected 175 CGs
down  to a limiting galaxy magnitude  of  $r^*=21$.  We have estimated
that  the median  redshift of  this   current version of  the  SDSS CG
catalog is $z_{\rm med} \approx 0.13$, substantially deeper than 
previous CG catalogs.  This catalog  will be useful for conducting
studies of the general characteristics of CGs, their environments, and
their component galaxies.

Our initial results show that SDSS CGs are, on average, about a factor
of  40 denser  than    their local surroundings;  that  their  general
physical   properties are similar   to those  of   other, less deep CG
catalogs;  that the fraction of  early-type galaxies is higher in SDSS
CGs than in the field;  and that there   is strong visual evidence  of
interactions and mergers within a significant fraction of SDSS CGs.

Our future goals are three-fold: further analyses of the properties of
the current version  of  the SDSS CG   catalog with its  175  CGs, the
enlargement  of the  sample of  SDSS  CGs   as more SDSS  data  become
available, and improvement  of CG selection  techniques.  Clearly, the
next step is to apply our algorithm to the current SDSS  Data Release.
At a detection rate of slightly better than 1 CG per square degree, we
expect that the final SDSS CG catalog, based upon the a completed SDSS
covering up to one-quarter  of the sky,  will contain on the  order of
5,000 -- 10,000 CGs.


\acknowledgments

We thank the  anonymous referee for the many  useful  comments on this
text.

Funding for the creation and distribution of the SDSS Archive has been
provided  by  the  Alfred   P.  Sloan  Foundation,  the  Participating
Institutions, the  National Aeronautics and  Space Administration, the
National  Science  Foundation,  the  U.S. Department  of  Energy,  the
Japanese Monbukagakusho, and the Max Planck Society. The SDSS Web site
is http://www.sdss.org/.

The SDSS is managed by the Astrophysical Research Consortium (ARC) for
the Participating Institutions. The Participating Institutions are The
University of Chicago, Fermilab, the Institute for Advanced Study, the
Japan Participation  Group, The  Johns Hopkins University,  Los Alamos
National  Laboratory, the  Max-Planck-Institute for  Astronomy (MPIA),
the  Max-Planck-Institute  for Astrophysics  (MPA),  New Mexico  State
University, University of Pittsburgh, Princeton University, the United
States Naval Observatory, and the University of Washington.

This  research has made   use of the NASA/IPAC Extragalactic  Database
(NED),  which  is operated by   the Jet Propulsion  Laboratory, at the
California Institute of Technology,  under contract with the  National
Aeronautics and Space Administration.



\clearpage

\clearpage

\clearpage

\clearpage

\clearpage

\clearpage

\clearpage

\clearpage

\clearpage

\clearpage

\clearpage

\clearpage
\begin{figure*}
\centering
\caption{Rest-frame $M_{g^*}-M_{r^*}$ color vs.\ spectroscopic redshift.
{\em  Black  dots:} field galaxy  sample.  {Red  dots:} SDSSCG  galaxy
sample.  (Note:  The SDSSCG  galaxy  sample  only contains  those  158
SDSSCG galaxies with known spectroscopic redshifts.)
\label{gr_vs_z}}
\end{figure*}

\begin{figure*}
\centering
\caption{$r^*$-band absolute magnitude $M(r^*) - 5\log (h)$ vs.\ rest-frame color.
{\em Black dots:}  field galaxy sample. {Red dots:} SDSSCG galaxy sample.  
(Note:  The SDSSCG galaxy sample only contains those 158 SDSSCG galaxies
with known spectroscopic redshifts.)
\label{mag_vs_color}}
\end{figure*}

\begin{figure*}
\centering
\caption{(Rest-frame) Color vs.\ (Rest-frame) Color.
{\em Black dots:}  field galaxy sample. {Red dots:} SDSSCG galaxy sample.  
(Note:  The SDSSCG galaxy sample only contains those 158 SDSSCG galaxies 
with known spectroscopic redshifts.)
\label{color_vs_color}}
\end{figure*}

\begin{figure*}
\centering
\caption{Atlas of SDSS Compact Groups is available at 
{\tt http://home.fnal.gov/$\sim$sallam/LeeCG/}}
\label{fig:cg}
\end{figure*}

\clearpage
{\tiny
\begin{deluxetable}{@{\,}c@{\,}c@{\,}c@{\,}c@{\,}c@{\,}c@{\,}c@{\,}c@{\,}c@{\,}c@{\,}c@{\,}c@{\,}c@{\,}c@{\,}c@{\,}}
\tabletypesize{\tiny}
\tablecaption{The SDSS Compact Group Properties. \label{tab_cg}}
\tablewidth{0pt}
\tablehead{
 \colhead{CG} & 
 \colhead{$\alpha_{\rm circ}$} & 
 \colhead{$\delta_{\rm circ}$} & 
 \colhead{$\theta_G$} & 
 \colhead{$\mu_G$} & 
 \colhead{$N$} & 
 \colhead{$\Delta r^*$} & 
 \colhead{$\theta_N/\theta_G$} & 
 \colhead{$z^{\rm bgm}_{\rm photo}$}&  
 \colhead{$z^{\rm cg}_{\rm spectro}$}&  
 \colhead{$N^{\rm cg}_{\rm spectro}$}&     
 \colhead{Comments\tablenotemark{*}}  \\
 \colhead{} &
 \colhead{[J2000.0]} &
 \colhead{[J2000.0]} &
 \colhead{[arcsec]} &
 \colhead{[mag arcsec$^{-2}$]} &
 \colhead{} &
 \colhead{} &
 \colhead{} &
 \colhead{} &
 \colhead{} &
 \colhead{} &
 \colhead{} &
 \colhead{} &
 \colhead{} \\
 \colhead{(1)} &
 \colhead{(2)} &
 \colhead{(3)} &
 \colhead{(4)} &
 \colhead{(5)} &
 \colhead{(6)} &
 \colhead{(7)} &
 \colhead{(8)} &
 \colhead{(9)} &
 \colhead{(10)} &
 \colhead{(11)} &
 \colhead{(12)} 
}
\startdata   
1& 09:40:48.72& -00:51:36.63& 38.9& 23.93& 4& 2.948& 3.64& 0.1290& \nodata& 0&  \\
2& 09:41:05.28& -00:28:19.98& 24.5& 22.91& 4& 2.073& 6.57& 0.1459& 0.1478& 1&  M(bc)\\
3& 09:49:58.08& -00:52:57.18& 28.7& 22.56& 4& 2.814& 8.33& 0.1398& 0.0916& 1&  \\
4& 09:56:09.36&  00:19:35.67& 18.1& 22.08& 4& 2.529& 15.46& 0.0056& 0.0353& 2&  M(ab)\\
5& 09:57:32.64&  00:40:03.81& 14.3& 21.77& 4& 0.876& 4.39& 0.0372& 0.0870& 1&  M(abc)\\
6& 10:00:24.96&  00:15:00.61& 20.2& 22.94& 5& 2.987& 3.49& 0.2374& 0.2195& 1&  I(ac)\\
7& 10:01:25.20&  01:07:32.08& 26.1& 23.15& 4& 2.825& 3.92& 0.0792& \nodata& 0&  \\
8& 10:01:26.88&  00:20:43.57& 26.2& 23.44& 5& 2.295& 4.23& 0.2228& 0.2316& 1&  \\
\enddata                                                     
\tablenotetext{*}{Endnotes:\\
                  +1:        a faint background galaxy is visible which could be a member of the group, \\
                  +2:        two faint background galaxies are visible which could be members of the group, \\
                  isolated?: faint background galaxies are visible which could be members of the group, \\
                  M(ab):     galaxies a and b appear to be merging, \\
                  I(ab):     galaxies a and b appear to be interacting, \\ 
                  T(a):      galaxy a shows a tidal tail, \\
                  h(abc):    galaxies a, b and c appear to be embedded within a common halo.\\
		  QSO(a):    galaxy a is classified as a QSO in NED}
\end{deluxetable} 
}                                                    

\begin{deluxetable}{@{\,}l@{\,}c@{\,}c@{\,}c@{\,}c@{\,}c@{\,}c@{\,}c@{\,}c@{\,}c@{\,}c@{\,}c@{\,}c@{\,}c@{\,}c@{\,}r@{\,}r@{\,}r@{\,}}                    
\rotate                    
\tabletypesize{\tiny}                    
\tablecaption{The SDSS Compact Group Members \label{tab_cgmem}}               
\tablewidth{0pt}                    
\tablehead{                    
\colhead{Name}&                    
\colhead{$\alpha$}&                    
\colhead{$\delta$} &                   
\colhead{$r^*$} &                   
\colhead{$\sigma_{r^*}$}&                    
\colhead{$A(r^*)$}&                    
\colhead{$u^*-g^*$} &                   
\colhead{$\sigma_{u^*-g^*}$}&                    
\colhead{$g^*-r^*$}&                    
\colhead{$\sigma_{g^*-r^*}$} &                   
\colhead{$r^*-i^*$} &                   
\colhead{$\sigma_{r^*-i^*}$} &                   
\colhead{$i^*-z^*$} &                   
\colhead{$\sigma_{i^*-z^*}$} &                   
\colhead{$z_{\rm sp}$} &                  
\colhead{$z_{\rm NED}$} &                  
\colhead{Notes} \\                   
\colhead{} &                   
\colhead{[J2000.0]} &                   
\colhead{[J2000.0]} &                   
\colhead{} &                   
\colhead{} &                   
\colhead{} &                   
\colhead{} &                   
\colhead{} &                   
\colhead{} &                   
\colhead{} &                   
\colhead{} &                   
\colhead{} &                   
\colhead{} &                   
\colhead{} &                   
\colhead{} &                   
\colhead{} &                   
\colhead{} &                   
\colhead{} \\                   
\colhead{(1)} &                   
\colhead{(2)} &                   
\colhead{(3)} &                   
\colhead{(4)} &                   
\colhead{(5)} &                   
\colhead{(6)} &                   
\colhead{(7)} &                   
\colhead{(8)} &                   
\colhead{(9)} &                   
\colhead{(10)} &                   
\colhead{(11)} &                   
\colhead{(12)} &                   
\colhead{(13)} &                   
\colhead{(14)} &                   
\colhead{(15)} &                   
\colhead{(16)} &                   
\colhead{(17)} \\                   
}                    
\startdata                    
1a& 09:40:49.77& -00:51:47.07& 16.707& 0.007& 0.108& 2.219& 0.093& 1.134& 0.010& 0.498& 0.008& 0.349& 0.008& \nodata& \nodata& \\    
1b& 09:40:49.80& -00:51:26.92& 18.123& 0.012& 0.108& 2.047& 0.231& 1.129& 0.022& 0.480& 0.015& 0.344& 0.024& \nodata& \nodata& \\    
1c& 09:40:47.40& -00:51:39.69& 18.487& 0.024& 0.107& 1.187& 0.138& 0.633& 0.034& 0.328& 0.032& 0.107& 0.073& \nodata& \nodata& \\    
1d& 09:40:48.99& -00:51:41.70& 19.655& 0.029& 0.108& 0.847& 0.133& 0.537& 0.043& 0.393& 0.041& 0.162& 0.091& \nodata& \nodata& \\    
& & & & & & & & & & & & & & & & \\ 
2a& 09:41:04.64& -00:28:13.50& 16.895& 0.008& 0.158& 2.072& 0.103& 1.155& 0.010& 0.461& 0.008& 0.382& 0.010& \nodata& \nodata& \\    
2b& 09:41:06.09& -00:28:24.50& 17.818& 0.012& 0.158& 1.488& 0.104& 1.009& 0.016& 0.545& 0.013& 0.403& 0.018& 0.1478& \nodata& \\    
2c& 09:41:05.90& -00:28:28.81& 18.309& 0.020& 0.157& 1.006& 0.106& 0.650& 0.026& 0.459& 0.025& 0.058& 0.055& \nodata& \nodata& \\
\enddata                    
\end{deluxetable}                    
   

\clearpage
\begin{deluxetable}{l r ll ll ll ll ll}
\tabletypesize{\tiny}
\tablecaption{Comparison between SDSS Compact Group Properties and other CGs \label{tab:Cgstat}}
\tablewidth{0pt}
\tablehead{
\colhead{Catalog}                                & 
\colhead{\# of CGs}                              &
\multicolumn{2}{c}{$N$}                          &
\multicolumn{2}{c}{$\mu_G$}                      &
\multicolumn{2}{c}{$z$}                          &
\multicolumn{2}{c}{$\theta_G$}                   &
\multicolumn{2}{c}{$D\times\theta_G$}            \\
\colhead{}                                       & 
\colhead{}                                       &
\multicolumn{2}{c}{}                             &
\multicolumn{2}{c}{[mag arcsec$^{-2}$]}          &
\multicolumn{2}{c}{}                             &
\multicolumn{2}{c}{[deg]}                        &
\multicolumn{2}{c}{[$h^{-1}$kpc]}                \\
\colhead{}                                       &
\colhead{}                                       &
\colhead{mean}                                   &
\colhead{median}                                 &
\colhead{mean}                                   &
\colhead{median}                                 &
\colhead{mean}                                   &
\colhead{median}                                 &
\colhead{mean}                                   &
\colhead{median}                                 &
\colhead{mean}                                   &
\colhead{median}                                 
}
\startdata
SDSSCG  &   175                    &   4.25$\pm$0.04  &  4.00   &   23.39$\pm$0.04                   &  23.51                      &   0.135$\pm$0.005\tablenotemark{a}   & 0.126\tablenotemark{a}        &   0.0088$\pm$0.0003  &  0.0073    &    58$\pm$2\tablenotemark{a}  &  55\tablenotemark{a}     \\
PCG     &    84                    &   4.15$\pm$0.04  &  4.00   &   23.25$\pm$0.05\tablenotemark{b}  &  23.32\tablenotemark{b}     &   \nodata                            & $\approx0.1$\tablenotemark{c} &   0.0114$\pm$0.0003  &  0.0113    &    \nodata                    & \nodata \\
SCG     &   121                    &   4.29$\pm$0.05  &  4.00   &   25.50$\pm$0.08\tablenotemark{d}  &  25.68\tablenotemark{d}     &   \nodata                            & \nodata                       &   0.15$\pm$0.01      &  0.125     &    \nodata                    & \nodata \\
UZC CG  &   291                    &   3.39$\pm$0.05  &  3.00   &   \nodata                          &  \nodata                    &   0.0173$\pm$0.0004                  & 0.0171                        &   \nodata            &  \nodata   &   147$\pm$3                   & 152     \\
LCCG    &    76                    &   3.33$\pm$0.08  &  3.00   &   \nodata                          &  \nodata                    &   0.079$\pm$0.003                    & 0.075                         &   0.016$\pm$0.001    &  0.013     &    53$\pm$2                   &  52     \\
RSCG    &    89                    &   3.65$\pm$0.14  &  3.00   &   \nodata                          &  \nodata                    &   0.014$\pm$0.001                    & 0.014                         &   0.18$\pm$0.02      &  0.08      &    60$\pm$3                   &  55     \\
HCG     &    92\tablenotemark{e}   &   4.55$\pm$0.09  &  4.00   &   23.4$\pm$0.1\tablenotemark{f}    &  23.4\tablenotemark{f}      &   0.034$\pm$0.002                    & 0.030                         &   0.061$\pm$0.005    &  0.048     &    84$\pm$5                   &  70     \\
\enddata
\tablenotetext{a}{Based upon the 131 SDSS CGs (75\%) that have a spectroscopic redshift determination}
\tablenotetext{b}{Converted from Thuan-Gunn $r_{\rm TG}$ to SDSS $r^*$ magnitudes assuming $r^*-r_{\rm TG} \approx 0.1$ for an elliptical galaxy; see \citet{Thuan76}, \citet{Smith02}, and \citet{Fukugita95}} 
\tablenotetext{c}{Determined photometrically, based upon the magnitude distribution of the brightest group members; see \citet{Iovino03}}
\tablenotetext{d}{Converted from $b_j$ to $r^*$ magnitudes assuming $b_j-r^* \approx 1.0$ for an elliptical galaxy; see \citet{Blanton01} and \citet{Iovino02}} 
\tablenotetext{e}{Here, we consider only the HCGs from the \citet{Hickson93} ``cleaned'' sample.}
\tablenotetext{f}{Converted from $E$ to $r^*$ magnitudes assuming $r^*-E \approx 0.3$ for an elliptical galaxy; see \citet{Blanton01} and \citet{Iovino02}} 
\end{deluxetable}

\clearpage
\begin{deluxetable}{lrrrr}
\tabletypesize{\small}
\tablecaption{Mean and Median Properties for the SDSSCG Galaxy and SDSS Field Galaxy Samples \label{tab:MeanMedianCompare}}
\tablewidth{0pt}
\tablehead{
\colhead{Test}  & 
\multicolumn{2}{c}{SDSSCG Sample}  & 
\multicolumn{2}{c}{SDSS Field Sample} \\
\colhead{}  & 
\colhead{mean}  & 
\colhead{median}  & 
\colhead{mean}  & 
\colhead{median}
}
\startdata
$z_{\rm sp}$ 	    &     0.130$\pm$0.005 &      0.120 &       0.130$\pm$0.002  &      0.120 \\
$M_{r^*}- 5\log (h)$& $-$21.166$\pm$0.096 &  $-$21.264 &   $-$21.292$\pm$0.028  &  $-$21.293 \\
$M_{u^*}-M_{g^*}$ &      1.550$\pm$0.025  &      1.659 &       1.485$\pm$0.007  &      1.575 \\
$M_{g^*}-M_{r^*}$ &      0.710$\pm$0.012  &      0.756 &       0.683$\pm$0.003  &      0.730 \\
$M_{r^*}-M_{i^*}$ &      0.364$\pm$0.008  &      0.393 &       0.351$\pm$0.003  &      0.378 \\
$M_{i^*}-M_{z^*}$ &      0.233$\pm$0.016  &      0.272 &       0.240$\pm$0.006  &      0.268 \\
\enddata
\end{deluxetable}

\clearpage
\begin{deluxetable}{lcc}
\tabletypesize{\small}
\tablecaption{Results from the 1D Kolmogorov-Smirnov Tests \label{tab:1DKS}}
\tablewidth{0pt}
\tablehead{
\colhead{Test}  & 
\colhead{KS Statistic}  & 
\colhead{Probability}
}
\startdata
$z_{\rm sp}$         &  0.01203  &  1.00000 \\
$M_{r^*}-5\log (h)$  &  0.06139  &  0.65114 \\
$M_{u^*}-M_{g^*}$    &  0.19557  &  0.00003 \\
$M_{g^*}-M_{r^*}$    &  0.16519  &  0.00079 \\
$M_{r^*}-M_{i^*}$    &  0.13418  &  0.01135 \\
$M_{i^*}-M_{z^*}$    &  0.05380  &  0.80009 \\
\enddata
\end{deluxetable}

\clearpage
\begin{deluxetable}{lcc}
\tabletypesize{\small}
\tablecaption{Results from the 2D Kolmogorov-Smirnov Tests \label{tab:2DKS}}
\tablewidth{0pt}
\tablehead{
\colhead{Test}  & 
\colhead{KS Statistic} &
\colhead{Probability} 
}
\startdata
$z_{\rm sp}$          vs.\ $M_{g^{*}}-M_{r^{*}}$    &  0.05316  &  0.90385 \\
                                                    &           &          \\
$M_{u^{*}}-M_{g^{*}}$ vs.\ $M_{r^{*}}-5\log (h)$    &  0.19937  &  0.00054 \\
$M_{g^{*}}-M_{r^{*}}$ vs.\ $M_{r^{*}}-5\log (h)$    &  0.17627  &  0.00296 \\
$M_{r^{*}}-M_{i^{*}}$ vs.\ $M_{r^{*}}-5\log (h)$    &  0.14778  &  0.02209 \\
$M_{i^{*}}-M_{z^{*}}$ vs.\ $M_{r^{*}}-5\log (h)$    &  0.07911  &  0.54591 \\
                                               &                &          \\
$M_{g^{*}}-M_{r^{*}}$ vs.\ $M_{u^{*}}-M_{g^{*}}$    &  0.23006  &  0.00002 \\
$M_{r^{*}}-M_{i^{*}}$ vs.\ $M_{g^{*}}-M_{r^{*}}$    &  0.18354  &  0.00109 \\ 
$M_{i^{*}}-M_{z^{*}}$ vs.\ $M_{r^{*}}-M_{i^{*}}$    &  0.15791  &  0.00971 \\
\enddata
\end{deluxetable}


\end{document}